\documentclass[twocolumn,showpacs,preprintnumbers,superscriptaddress,
               amsmath,amssymb,prd]{revtex4}
\usepackage[dvips]{graphicx}

\newcommand{\txt}{\textstyle}

\newcommand{\half} {{\txt \frac{1}{2}}}
\newcommand{\third}{{\txt \frac{1}{3}}}
\newcommand{\twothirds}{{\txt \frac{2}{3}}}
\newcommand{\tr}{\text{tr}}

\newcommand{\feyn}[1]{
  \setbox0=\hbox{\ensuremath{#1}}
  \hbox to\wd0{\hbox to0pt{\hbox to\wd0{\hss/\hss}\hss}\box0}}

\newcommand{\MeV}{\;{\rm MeV}}
\newcommand{\diag}{\text{diag}}
\newcommand{\Qtilde}{{\tilde Q}}
\newcommand{\mue}{\mu_e}
\newcommand{\ms}{M_s}
\newcommand{\mssq}{\ms^2/\mu}
\newcommand{\energy}[2]{\varepsilon_{\text{$#1$-$#2$}}}
\newcommand{\aenergy}[2]{\tilde{\varepsilon}_{\text{$#1$-$#2$}}}
\newcommand{\mubar}[2]{\bar{\mu}_{\text{$#1$-$#2$}}}
\newcommand{\deltamu}[2]{\delta\mu_{\text{$#1$-$#2$}}}
\newcommand{\Pidiag}[2]{\Pi^{#1}_{\text{diag($#2$)}}}
\newcommand{\Pioff}[5]{\Pi^{#1}_{\text{off($#2$-$#3$;$#4$-$#5$)}}}
\newcommand{\fph}{f_{\text{p-p}}}
\newcommand{\fpa}{f_{\text{p-a}}}

\begin{document}

\preprint{RBRC-524}

\title{Analytical and numerical evaluation of the Debye and Meissner
 masses \\ in dense neutral three-flavor quark matter}

\author{Kenji Fukushima}
\affiliation{RIKEN BNL Research Center, Brookhaven National
 Laboratory, Upton, New York 11973, USA}


\begin{abstract}
We calculate the Debye and Meissner masses and investigate
chromomagnetic instability associated with the gapless color
superconducting phase changing the strange quark mass $\ms$ and the
temperature $T$.  Based on the analytical study, we develop a
computational procedure to derive the screening masses numerically
from curvatures of the thermodynamic potential.  When the temperature
is zero, from our numerical results for the Meissner masses, we find
that instability occurs for $A_1$ and $A_2$ gluons entirely in the
gapless color-flavor locked (gCFL) phase, while the Meissner masses
are real for $A_4$, $A_5$, $A_6$, and $A_7$ until $\ms$ exceeds a
certain value that is larger than the gCFL onset.  We then handle
mixing between color-diagonal gluons $A_3$, $A_8$, and photon
$A_\gamma$, and clarify that, among three eigenvalues of the mass
squared matrix, one remains positive, one is always zero because of an
unbroken ${\rm U}(1)_\Qtilde$ symmetry, and one exhibits
chromomagnetic instability in the gCFL region.  We also examine the
temperature effects that bring modifications into the Meissner
masses. The instability found at large $\ms$ for $A_4$, $A_5$, $A_6$,
and $A_7$ persists at finite $T$ into the $u$-quark color
superconducting (uSC) phase which has $u$-$d$ and $s$-$u$ but no
$d$-$s$ quark pairing and also into the two-flavor color
superconducting (2SC) phase characterized by $u$-$d$ quark pairing
only.  The $A_1$ and $A_2$ instability also goes into the uSC phase,
but the 2SC phase has no instability for $A_1$, $A_2$, and $A_3$.  We
map the unstable region for each gluon onto the phase diagram as a
function of $\ms$ and $T$.
\end{abstract}


\maketitle


\section{Introduction}

The rich phase structure of matter at high baryon density has
attracted much interest over decades theoretically and
phenomenologically.  At sufficiently high density and low temperature,
wherever quarks feel an attractive force as is suggested from the one
gluon exchange between quarks that are antisymmetric in color, the
Fermi surface of quark matter is unstable against forming  a Cooper
pair whose condensation leads to color
superconductivity~\cite{reviews}.  It is well established that the
color-flavor locked (CFL) phase is a ground state of three-flavor
quark matter in the asymptotically high density region where the
strange quark mass $\ms$ is negligible as compared with the quark
chemical potential $\mu$~\cite{Alford:1998mk}.

If it is realized in a bulk system like the cores of compact stellar
objects, quark matter must be neutral in electric and color
charges~\cite{Iida:2000ha,Alford:2002kj,Steiner:2002gx}.  As long as
$\ms/\mu$ is negligibly small, three-flavor quark matter, which is
composed of the equal number of $u$, $d$, and $s$ quarks, satisfies
electric and color neutrality on its own regardless of whether it is
in the normal or CFL phase.  When $\ms$ comes to have a substantial
effect suppressing $s$ quarks, there arise various possibilities in
forming Cooper pairs and in achieving neutrality, that brings
intricate subtleties into the phase structure especially in the
intermediate density region where $\ms$ can compete $\mu$.

When the system is normal quark matter, a finite electron density is
required to neutralize the system electrically because the number of
$s$ quarks is reduced by $\ms\neq0$.  In CFL quark matter, on the
other hand, neutrality can be fulfilled even without electrons rigidly
at zero temperature~\cite{Rajagopal:2000ff} and approximately at low
temperatures below an insulator-metal
crossover~\cite{Ruster:2004eg,Fukushima:2004zq}.  This is because the
BCS ansatz for Cooper pairing between cross-species of quarks enforces
equality in the numbers of red, green, blue, and $u$, $d$, $s$ quarks
and thus neutrality.  At zero temperature, another candidate, that is,
the two-flavor color superconducting (2SC) phase, is known to cost a
larger free energy than the CFL phase under the neutrality conditions,
or, it could exist for strong coupling between quarks in some density
windows~\cite{Alford:2002kj,Alford:2003fq,Fukushima:2004zq}.

Electronless and neutral CFL quark  matter lasts as long as the Fermi
energy mismatch when two quarks were not paired, which is given by
$\ms^2/2\mu$ as explained later, is less than the energy gap
$\Delta$.  There should appear a new state of matter once the energy
gain by releasing the pressure between cross-species of quarks
surpasses the gap energy, which occurs when $\ms^2/2\mu>\Delta$.
Then, some of quark energy dispersion relations pass across zero and
their pairing is disrupted in the corresponding momentum (blocking)
region that is from one zero to another zero of the quark excitation
energy.  The new state is called the gapless CFL (gCFL)
phase~\cite{Alford:2003fq} and, if any other phase transitions such as
the chiral phase transition and the transition to a crystalline color
superconductor~\cite{loff,Alford:2000ze} lie at lower densities than
the gapless onset, the gCFL phase must show up next as the density
goes down from the CFL side.

The gapless or breached pairing superconductivity was first discussed
in a non-relativistic model and was recognized as an unstable
state~\cite{Sarma} and its analogue in three-flavor quark matter was
considered in Ref.~\cite{Alford:1999xc}.  It was discovered later on
that stable gapless superconductivity could be possible under the
constraint of particle number fixing in a non-relativistic
model~\cite{Gubankova:2003uj} (see also Ref.~\cite{Forbes:2004cr} for
more discussions on stability) and of electric and color neutrality in
the gapless regime of the 2SC (g2SC) phase~\cite{Shovkovy:2003uu}.

Besides gapless superconductivity, there is another competing
possibility, i.e.\ the mixed phase, to realize
neutrality~\cite{Neumann:2002jm,Shovkovy:2003ce,Bedaque:2003hi}, and
if the mixed phase has a lower free energy, the gapless phase would
not come to realization.  According to Ref.~\cite{Reddy:2004my} it can
be the case in fact for the g2SC phase which should be taken over
energetically by the mixed phase where normal and color
superconducting phases coexist, and consequently the g2SC phase may be
less of a reality.  It must deserve further investigation, however,
especially with the screening effects in the mixed phase taken into
account to declare the existence or non-existence of the (g)2SC phase.
In the case of the gCFL phase, on the other hand, the free energy
comparison indicates that the gCFL phase should be stable
energetically apart from a minor exception of the CFL-2SC mixed phase
possibility, which will be presumably ruled out, however, once the
surface tension and the Coulomb energy corrections are
included~\cite{Alford:2004nf}.  Hence, so far, the gCFL phase remains
as a likely candidate for the ground state at moderate density and is
considered to be relevant to neutron star
physics~\cite{Alford:2004zr}.

However, chromomagnetic instability first found in the g2SC
phase~\cite{Huang:2004bg} and later confirmed also in the gCFL
phase~\cite{Casalbuoni:2004tb,Alford:2005qw} implies that the gapless
phase is not stable against perturbation of transverse
(chromomagnetic) gluon fields and the true ground state must need
something further.  It is negative Meissner masses squared, i.e.,
imaginary Meissner masses, that the authors of
Refs.~\cite{Huang:2004bg,Casalbuoni:2004tb,Alford:2005qw} have
actually revealed in the gapless phases.  Since the Meissner mass is
the screening mass for transverse gluons, the instability signifies
spontaneous generation of the expectation value of gauge fields.
Keeping in mind a simple relation in superconductors between the gauge
fields and currents in the London gauge, we can regard chromomagnetic
instability as color current generation.  (An interpretation of
instability as spontaneous \textit{baryon} current generation has been
argued in Ref.~\cite{Huang:2005pv}.)  To put it another way, it is
possible to interpret it as instability towards a state characterized
by diquark condensates with color-dependent oscillation in
coordinate space~\cite{Giannakis:2004pf}, because gauge fields in the
quark propagator generally reside in the form of the covariant
derivative.  This state of quark matter is a sort of crystalline color
superconductors under the single \textit{colored} plain wave ansatz.
It should be worth emphasizing that the difference is only in
intuitive pictures, and in effect, these above interpretations are
equivalent, which is mathematically described by the gauge
transformation.  Still, anyway, it is controversial what the true
ground state should be that supersedes the homogeneous gCFL phase.

\begin{figure}
\includegraphics[width=7cm]{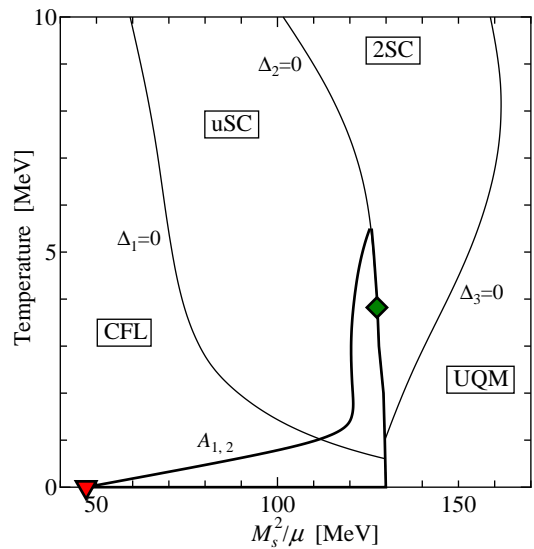}
\caption{Enclosed with solid thick lines is the unstable region for
$A_{1,2}$ on the phase diagram (see Fig.~1 in
Ref.~\cite{Fukushima:2004zq} and explanations therein).  The critical
end-point is marked by a square below which the phase transition from
the uSC to the 2SC phase or unpaired quark matter (UQM) is first
order.  A triangle indicates the gCFL onset located at
$\mssq=47.1\MeV$.}
\label{fig:phase12}
\end{figure}

\begin{figure}
\includegraphics[width=7cm]{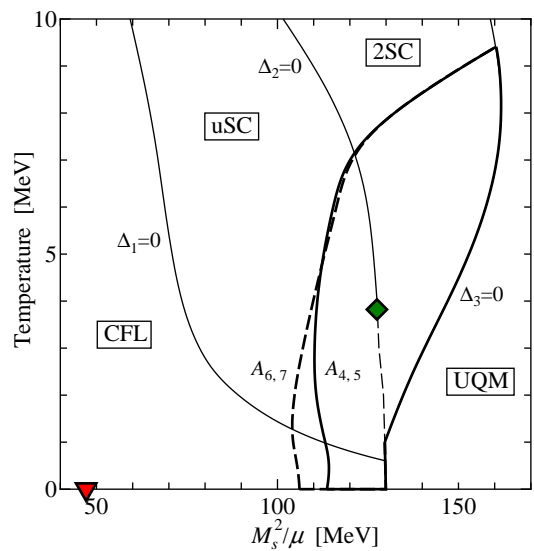}
\caption{Enclosed with solid and dashed thick lines are the unstable
regions for $A_{4,5}$ and $A_{6,7}$ respectively.}
\label{fig:phase4567}
\end{figure}

\begin{figure}
\includegraphics[width=7cm]{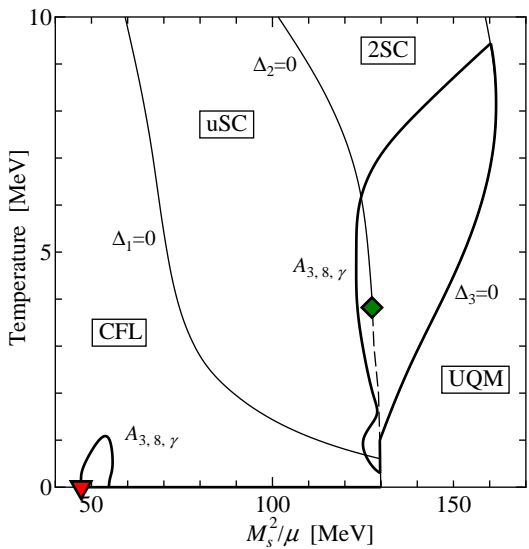}
\caption{Enclosed with solid thick lines is the unstable region for
one or two of eigenmodes consisting of $A_3$, $A_8$, and $A_\gamma$.}
\label{fig:phase380}
\end{figure}

The purpose of this paper is to compute the Debye and Meissner masses
in the (g)CFL phase and to present detailed analyses on the nature of
chromomagnetic instability which occurs in connection with the
presence of gapless quarks.  Our central results are condensed in the
unstable regions on the phase diagram presented in
Figs.~\ref{fig:phase12}, \ref{fig:phase4567}, and \ref{fig:phase380}.
In both the CFL and gCFL phases there is no mixing with other gluons
for $A_1$, $A_2$, $A_4$, $A_5$, $A_6$, and $A_7$, among which two
gluons of $(A_1,A_2)$, $(A_4,A_5)$, and $(A_6,A_7)$ respectively have
the same screening masses.  We shall write $A_{1,2}$ for instance to
mean whichever of $A_1$ or $A_2$ equivalently from now on.  As for
$A_3$, $A_8$, and the electromagnetic field $A_\gamma$, on the other
hand, mixing between them arises and thus we have to refer to
eigenvalues of the $3\times3$ mass squared matrix to investigate
instability.  It should be noted that the color indices are labeled
according to the conventional Gell-Mann matrices in color space.

Figure \ref{fig:phase12} shows the unstable region enclosed with thick
lines inside which the Meissner mass squared for $A_{1,2}$ is
negative.  The phase diagram underlaid is what has been revealed in
Ref.~\cite{Fukushima:2004zq} for the case of weak coupling yielding
$\Delta_0=25\MeV$ at $\ms=T=0$.  The gCFL phase starts appearing at
$\mssq=47.1\MeV$ which is well close to the kinematical estimate
$(\mssq)_{\rm c}\simeq 2\Delta$.  In the vicinity of the phase
boundary line on which one of three gaps vanishes, i.e., $\Delta_2=0$,
there comes out a rather complicated structure which is not robust but
strongly depends on a choice of the coupling strength.  We will not
present results for other coupling cases as done in
Ref.~\cite{Fukushima:2004zq}, for our purpose here is not to disclose
the phase structure but to argue characteristic features of
chromomagnetic instability and to settle where it becomes relevant on
the phase diagram.

We observe that instability grows for $A_{1,2}$ as soon as the gCFL
state occurs at zero temperature.  The instability penetrates from the
CFL region where all $u$-$d$, $d$-$s$, and $s$-$u$ quarks make Cooper
pairs into the uSC region where $u$-$d$ and $s$-$u$ quarks remain to
pair, but never enters the 2SC region where only the $u$-$d$ quark
pairing gap is nonvanishing.  Actually in the 2SC phase, $A_{1,2}$
become irrelevant to the Higgs-Anderson mechanism, and therefore the
Meissner mass for $A_{1,2}$ must be zero there.

The unstable regions for $A_{4,5}$ and $A_{6,7}$ are depicted in
Fig.~\ref{fig:phase4567}.  In the 2SC phase there is no discrimination
between $A_{4,5}$ and $A_{6,7}$ and the difference is manifested only
when quark matter lies in the CFL or uSC phase.  Our mapping is
consistent with all of the known results both in the 2SC
phase~\cite{Huang:2004bg} and in the CFL
phase~\cite{Casalbuoni:2004tb};  no instability takes place near the
gCFL onset for $A_{4,5}$ nor $A_{6,7}$, while the large $\mssq$
regions exhibit instability which turns out to be related to the g2SC
instability.  At small temperatures and large $\mssq$ the Meissner
masses squared for $A_{4,5}$ and $A_{6,7}$ are negative until the
system reaches the phase boundary from which the system enters the
phase of unpaired quark matter (UQM).

It seems to be somewhat tricky to understand the unstable regions for
mixed modes of $A_3$, $A_8$, and $A_\gamma$ shown in
Fig.~\ref{fig:phase380}.  At zero or extremely low temperatures whose
energy scale is determined by the electron contribution as discussed
later, the whole gCFL phase is unstable and our results qualitatively
agree with Ref.~\cite{Casalbuoni:2004tb}.  At finite temperature there
are two distinct regions where instability remains.  The instability
near the gCFL onset eventually disappears as the temperature grows,
while the instability at larger $\mssq$ goes further into the 2SC
phase.  In fact in the 2SC phase, $A_3$ decouples from others and its
Meissner mass is reduced to zero, and one of the two eigenmodes
composed of $A_8$ and $A_\gamma$ exhibits chromomagnetic instability
as is consistent with the findings in the g2SC
phase~\cite{Huang:2004bg}.  For even stronger coupling when the 2SC
(not necessarily g2SC) phase is possible at zero temperature (see
Fig.~17 in Ref.~\cite{Fukushima:2004zq}), we have numerically checked
that instability at larger $\mssq$ starts exactly at the g2SC onset.

The exact correspondence between the gapless and instability onsets is
apparent only at zero temperature since the finite-temperature
effects allow for thermal quark excitations which are not clearly
distinguishable from quarks in the blocking region.  An interesting
manifestation of this is the unstable region in the uSC phase in
which there are no gapless quarks below $T\sim 5\MeV$ (see the dotted
curve in the uSC region shown in Fig.~1 in
Ref.~\cite{Fukushima:2004zq}).  In a sense, at finite temperature, we
can say that the instability penetrates into gapped sides, as is also
observed from Fig.~5 in Ref.~\cite{Alford:2005qw}.

We are explaining how we have come by these instability mappings in a
numerical way in later discussions, starting with the following
subsections in which we shall make a brief review of the gCFL phase
and of the derivation of the Debye and Meissner screening masses of
gauge fields.  The readers who are already familiar with these basics
can skip most of them and jump to Sec.~\ref{sec:self-energies}.


\subsection{Gapless CFL phase}

Our strategy to approach the Debye and Meissner masses is based on the
thermodynamic potential that has been formulated within an effective
model.  In the following subsubsections, we discuss the model and
approximations to describe the gCFL phase, and then illustrate the
quark excitation energies as a function of the momentum (i.e.\ the
dispersion relations).


\subsubsection{Model and approximations}

In this paper we adopt essentially the same model and approximations
as used in Refs.~\cite{Alford:2003fq,Fukushima:2004zq}.  The only
difference is that the $\ms$ effect is incorporated as an effective
chemical potential shift.  This approximation is necessary to relate
the Meissner mass to the potential curvature with respect to gluon
source fields in a simple way.  The model employed here is the
Nambu--Jona-Lasinio (NJL) model with four-fermion interaction.  In our
model study we assume that the predominant diquark condensate is
antisymmetric in Dirac indices, antisymmetric in color, and thus
antisymmetric in flavor;
\begin{equation}
 \langle \psi^a_i C\gamma_5 \psi^b_j \rangle \sim
  \Delta_1 \epsilon^{ab 1}\epsilon_{ij1} \!+\!
  \Delta_2 \epsilon^{ab 2}\epsilon_{ij2} \!+\!
  \Delta_3 \epsilon^{ab 3}\epsilon_{ij3}\,,
\label{eq:diquark}
\end{equation}
where ($i$,$j$) and ($a$,$b$) represent the flavor indices
($u$,$d$,$s$) and the color triplet indices (red,green,blue)
respectively.  The gap parameters $\Delta_1$, $\Delta_2$, and
$\Delta_3$ describe a  $9\times 9$ matrix in color-flavor space that
takes the form,
\begin{equation}
 \boldsymbol{\Delta}=
\newcommand{\mDe}{\makebox[1.2em][r]{$-\Delta$}} 
 \left( \begin{array}{ccccccccc}
  0 & \Delta_3 & \Delta_2 & 0 & 0 & 0 & 0 & 0 & 0 \\
 \Delta_3 & 0 & \Delta_1 & 0 & 0 & 0 & 0 & 0 & 0 \\
 \Delta_2 & \Delta_1 & 0  & 0 & 0 & 0 & 0 & 0 & 0 \\
  0 & 0 & 0 & 0 & \mDe_3 & 0 & 0 & 0 & 0 \\
  0 & 0 & 0 & \mDe_3 & 0 & 0 & 0 & 0 & 0 \\
  0 & 0 & 0 & 0 & 0 & 0 & \mDe_2 & 0 & 0 \\
  0 & 0 & 0 & 0 & 0 & \mDe_2 & 0 & 0 & 0 \\
  0 & 0 & 0 & 0 & 0 & 0 & 0 & 0 & \mDe_1 \\
  0 & 0 & 0 & 0 & 0 & 0 & 0 & \mDe_1 & 0 \\
 \end{array} \right)
\label{eq:gap_matrix}
\end{equation}
in the basis $(ru,gd,bs, gu,rd, rs,bu,bd,gs)$.  In the same way as in
Refs.~\cite{Alford:2003fq,Fukushima:2004zq}, we ignore diquark
condensates which are symmetric in color, though we know that they do
not break any new symmetry and consequently can take a finite value.
This approximation is motivated by the fact that the QCD interaction
is repulsive between quarks in this channel and such diquark
condensates have been actually known to be insignificant
quantitatively~\cite{Ruster:2004eg}.

In the mean-field approximation where the condensates
(\ref{eq:diquark}) are relevant, we only have to consider the
four-fermion interaction in the diquark-diquark channel that can be
generally in hand after appropriate Fierz transformation.  That is,
\begin{equation}
 {\cal L} = \bar{\psi}(i\feyn{\partial}+\boldsymbol{\mu}\gamma^0
  -\boldsymbol{M})\psi + {\cal L}_{\text{int}} \,,
\label{eq:lagrangian}
\end{equation}
where
\begin{equation}
\begin{split}
 {\cal L}_{\text{int}} &= \frac{G}{4}(\bar{\psi}^a_i i\gamma_5
 \epsilon^{ab\eta}\epsilon_{ij\eta}C\bar{\psi}^{T b}_j) \\
  & \qquad\qquad\times  (\psi^{T a'}_{i'}C i\gamma_5
  \epsilon^{a'b'\eta}\epsilon_{i'j'\eta}
  \psi^{b'}_{j'}) \,.
\end{split}
\end{equation}
Here the mass matrix $\boldsymbol{M}$ is unity in color and
$\diag(0,0,\ms)$ in flavor ($u$,$d$,$s$) space in our
approximation and $\boldsymbol{\mu}$ the matrix of quark chemical
potentials in color-flavor space.  Bold symbols generally denote
matrices in color-flavor space.  The charge conjugation matrix is
$C=i\gamma^2\gamma^0$.

The effect of nonzero $\boldsymbol{M}$ on the particle dispersion
relation becomes apparent in the vicinity of the Fermi surface, where
it can be well approximated by a shift in quark chemical potentials by
$\boldsymbol{MM}^\dagger/2\mu$.  We shall make use of this
prescription that actually turns out to be essential to allow us to
relate the Meissner mass to the potential curvature.  In the present
work, as in Refs.~\cite{Alford:2003fq,Fukushima:2004zq},
$\boldsymbol{M}$ is not current but constituent quark mass and is
treated as an input parameter.  We know that this approximation
provides a correct phase structure unless the chiral phase transition
cuts deeply into the gCFL region under the choice of strong coupling
in the chiral sector~\cite{Abuki:2004zk}.

Chemical potentials are fixed by equilibration and neutrality.  Stable
quark matter should be a color singlet as a whole and neutral in the
electric charge under $\beta$-equilibrium.  Color singletness is a
more stringent condition than neutrality.  It has been shown, however,
that no energy cost is needed to project a neutral state onto a
singlet state in color in the thermodynamic limit~\cite{Amore:2001uf}.
Hence, it is sufficient to impose global neutrality with respect to
the electric and color charges.  To describe that, we shall consider
$\mue$, $\mu_3$, and $\mu_8$ that are coupled to \textit{negative}
$Q=\diag(\twothirds,-\third,-\third)$ in flavor ($u$,$d$,$s$) space so
that a positive $\mue$ corresponds to the electron density,
$T_3=\diag(\half,-\half,0)$, and
${\txt\frac{2}{\sqrt{3}}}T_8=\diag(\third,\third,-\twothirds)$ in
color (red,\,green,\,blue) space, respectively.  Then, 9 diagonal
components of the \textit{effective} chemical potential matrix
$\boldsymbol{\mu}_{\text{eff}}$ (including the shift by
$\boldsymbol{MM}^\dagger/2\mu$) in color-flavor space are explicitly
\begin{equation}
 \begin{array}{rcl}
 \mu_{ru}&=& \mu-\twothirds\mu_e+\half\mu_3+\third\mu_8,\\[1ex]
 \mu_{gd}&=& \mu+\third\mu_e-\half\mu_3+\third\mu_8,\\[1ex]
 \mu_{bs}&=& \mu+\third\mu_e-\twothirds\mu_8
  - {\txt \frac{\ms^2}{2\mu}},\\[3ex]

 \mu_{gu}&=& \mu-\twothirds\mu_e-\half\mu_3+\third\mu_8,\\[1ex]
 \mu_{rd}&=& \mu+\third\mu_e+\half\mu_3+\third\mu_8,\\[3ex]

 \mu_{rs}&=& \mu+\third\mu_e+\half\mu_3+\third\mu_8
  - {\txt \frac{\ms^2}{2\mu}},\\[1ex]
 \mu_{bu}&=& \mu-\twothirds\mu_e-\twothirds\mu_8,\\[3ex]

 \mu_{bd}&=& \mu+\third\mu_e-\twothirds\mu_8\ . \\[1ex]
 \mu_{gs}&=& \mu+\third\mu_e-\half\mu_3+\third\mu_8
  - {\txt \frac{\ms^2}{2\mu}},\\[1ex]
\end{array}
\label{eq:mu}
\end{equation}

The gap parameters corresponding to (\ref{eq:diquark}) in this model
are precisely defined as
\begin{equation}
 \Delta_\eta = \half G\langle \psi^{T\alpha}_i C i\gamma_5
  \epsilon^{\alpha\beta\eta}\epsilon_{ij\eta}\psi^\beta_j \rangle.
\end{equation}
With all these definitions, if we choose Nambu-Gor'kov basis as
$\Psi(p)=(\psi(p),C\bar{\psi}^T(-p))^T$, we can express the mean-field
(inverse) quark propagator in a simple form,
\begin{equation}
 i S^{-1}(p) = \left( \begin{array}{cc}
  \feyn{p} + \boldsymbol{\mu}\gamma^0 & i\gamma_5\boldsymbol{\Delta}\\
  i\gamma_5\boldsymbol{\Delta} & \feyn{p} - \boldsymbol{\mu}\gamma^0
 \end{array} \right)
\end{equation}
after rearranging the Dirac matrices.  The quark propagator is a
$72\times72$ matrix in color, flavor, spin, and Nambu-Gor'kov space.
From zeros of the inverse propagator, we can read 72 energy dispersion
relations $\varepsilon_i(p)$.  The thermodynamic potential $\Omega$ is
thus written in terms of $\varepsilon_i$'s as
\begin{equation}
\begin{split}
 \Omega &= -\frac{1}{8\pi^2}\int_0^\Lambda \!\! dp\,p^2
  \sum_{j=1}^{72} \Bigl\{|\varepsilon_i| + 2T\ln
  \bigl(1+e^{-|\varepsilon_i|/T}\bigr)\Bigr\} \\
 & +\frac{1}{G}\bigl(\Delta_1^2+\Delta_2^2+\Delta_3^2\bigr)
  -\frac{\mue^4}{12\pi^2}-\frac{\mue^2 T^2}{6}
  -\frac{7\pi^2 T^4}{180}
\end{split}
\label{eq:potential}
\end{equation}
with the electron contribution in the last three terms.  In order to
fix three gap parameters and three chemical potentials at each $\ms$,
$\mu$, and $T$, we will simultaneously solve three gap equations,
\begin{equation}
 \frac{\partial\Omega}{\partial\Delta_1}
 =\frac{\partial\Omega}{\partial\Delta_2}
 =\frac{\partial\Omega}{\partial\Delta_3}=0 \,,
\end{equation}
and three neutrality conditions,
\begin{equation}
 \frac{\partial\Omega}{\partial\mue}
 =\frac{\partial\Omega}{\partial\mu_3}
 =\frac{\partial\Omega}{\partial\mu_8}=0 \,.
\label{eq:neutrality}
\end{equation}
Here $\Lambda$ in (\ref{eq:potential}) is the ultraviolet cut-off
parameter and we use $\Lambda=800\MeV$ as in
Refs.~\cite{Alford:2003fq,Fukushima:2004zq}.  The coupling constant
$G$ is chosen to yield $\Delta_0=25\MeV$ when both $\ms$ and $T$ are
zero.  In the present work we take the quark chemical potential
$\mu=500\MeV$ that roughly corresponds to 10 times the normal nuclear
density in the model.

When $\ms$ is not so large as to disrupt any Cooper pair, $\mue$,
$\mu_3$, and $\mu_8$ satisfying electric and color neutrality have
been analyzed at zero temperature in a model-independent
way~\cite{Alford:2002kj}.  Two of three can be fixed as a function of
the third that we will choose $\mue$ here, then
\begin{align}
 \mu_3 &= \mue,
\label{eq:CFL_mu3} \\
 \mu_8 &= -\frac{\ms^2}{2\mu}+\frac{\mue}{2}.
\label{eq:CFL_mu8}
\end{align}
The important feature of the CFL phase we should note is that the
thermodynamic potential (\ref{eq:potential}) at $T=0$ with the
electron contribution disregarded is independent of
\begin{equation}
 \mu_{\Qtilde} = -{\txt \frac{4}{9}}\bigl(\mue
  +\mu_3 + \half\mu_8\bigr) ,
\label{eq:muQtilde}
\end{equation}
that is the chemical potential for
$\Qtilde=Q-T_3-\frac{1}{\sqrt{3}}T_8$.  It is known that $\Qtilde$ is
a generator of the ``rotated electromagnetism'' that is never broken
by any condensate of the form (\ref{eq:diquark}).  At zero temperature
$\Qtilde$-charged excitations in CFL quark matter are all gapped and
the CFL phase is a $\Qtilde$-insulator~\cite{Rajagopal:2000ff}.  

In the presence of the electron contribution to the thermodynamic
potential, in reality, there is a gentle curvature on the plateau of
the thermodynamic potential provided by the last three terms in
(\ref{eq:potential}) which select $\mu_e=0$, meaning zero electron
density.  Although it makes only slight changes in energy, the
existence of the electron effects is crucial in understanding the
$A_{1,2}$ instability.


\subsubsection{Gapless dispersion relations}

Since the gap matrix (\ref{eq:gap_matrix}) is block-diagonal, the
$72\times72$ quark propagator matrix is block-diagonal as well in
color-flavor space and can be divided into four parts;  one
$24\times24$ part for $ru$-$gd$-$bs$ quark pairing with $\Delta_1$,
$\Delta_2$, and $\Delta_3$ and three $16\times16$ parts for quark
pairing of $bd$-$gs$ with $\Delta_1$, $rs$-$bu$ with $\Delta_2$, and
$gu$-$rd$ with $\Delta_3$.  It might be possible but quite hard to
handle the $24\times24$ part analytically.  Fortunately, as we will
see shortly, this $24\times24$ intricate part has little to do with
chromomagnetic instability, so we will limit our discussion here to
the propagator for $bd$-$gs$, $rs$-$bu$, and $gu$-$rd$ pairings.

Leaving the explicit expression of the quark propagator until
Sec.~\ref{sec:self-energies}, let us elucidate here the quark energy
dispersion relations obtained from zeros of the inverse propagator.
In general for the $16\times16$ part involving two species $A$ and $B$
quarks the energy dispersion relation takes the form,
\begin{align}
 \energy{A}{B} &= \sqrt{(p-\mubar{A}{B})^2 +\Delta_{AB}^2} +
  \deltamu{A}{B} ,\\
 \aenergy{A}{B} &= \sqrt{(p+\mubar{A}{B})^2 +\Delta_{AB}^2} +
  \deltamu{A}{B}
\label{eq:dispersion}
\end{align}
for quasi-particle and antiparticle excitations respectively, where
$\mubar{A}{B}=\half(\mu_A+\mu_B)$ and
$\deltamu{A}{B}=\half(\mu_A-\mu_B)$ and $\Delta_{AB}$ is the energy
gap for $A$-$B$ pairing.  In our definition
$\deltamu{A}{B}=-\deltamu{B}{A}$.  Gapless dispersion relations come
about once
\begin{equation}
 |\deltamu{A}{B}| > \Delta_{AB}.
\label{eq:general}
\end{equation}

\begin{figure}
\includegraphics[width=7cm]{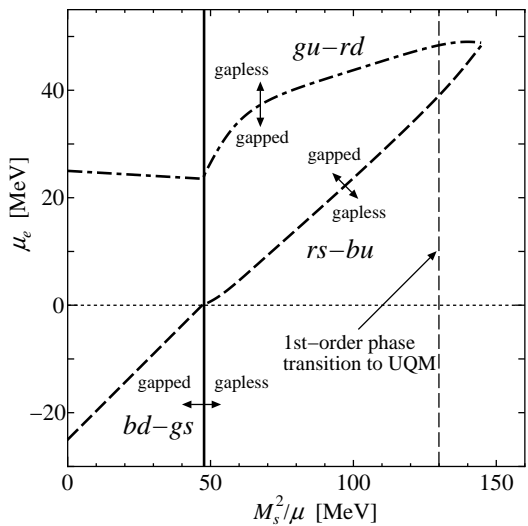}
\caption{Boundaries for the gapless conditions (\ref{eq:general}) for
$bd$-$gs$ (solid), $rs$-$bu$ (dashed), and $gu$-$rd$ (dot-dashed)
quarks.  Because $bd$-$gs$ quarks are neutral in $\Qtilde$-charge, the
system can go through the solid line, though the bandgap is edged with
the dashed and dot-dashed lines and quark matter is a
$\Qtilde$-insulator between these two boundaries.  With the inclusion
of the electron contributions to the thermodynamic potential, $\mue$
is zero until $\mssq$ reaches the $bd$-$gs$ line, and then $\mue$ goes
up slightly below the $rs$-$bu$ dashed boundary.  This is because the
$\Qtilde$-charged electrons must be canceled by a small disruption of
$rs$-$bu$ quark pairing.}
\label{fig:bandgap}
\end{figure}

Then, using effective chemical potentials (\ref{eq:mu}) with the known
CFL solutions (\ref{eq:CFL_mu3}) and (\ref{eq:CFL_mu8}), we have
\begin{align}
 \deltamu{gu}{rd} &= -\half\mue-\half\mu_3 = -\mue \,,\\
 \deltamu{rs}{bu} &= \half\mue+{\txt\frac{1}{4}}\mu_3+\half\mu_8
  -{\txt\frac{\ms^2}{4\mu}} = \mue-{\txt\frac{\ms^2}{2\mu}} \,,\\
 \deltamu{bd}{gs} &= {\txt\frac{1}{4}}\mu_3-\half\mu_8
  +{\txt\frac{\ms^2}{4\mu}} = {\txt\frac{\ms^2}{2\mu}} \,,
\end{align}
and
\begin{equation}
 \mubar{gu}{rd}=\mubar{bd}{gs}=\mubar{rs}{bu}
  = \mu - {\txt \frac{\ms^2}{6\mu}}
\end{equation}
in the CFL phase.  If it were not for the electron terms in the
thermodynamic potential, $\mue$ could lie anywhere as far as it does
not disrupt the pairing of $gu$-$rd$ nor $rs$-$bu$ quarks which are
$\Qtilde$-charged.  In the region between the dashed and dot-dashed
curves shown in Fig.~\ref{fig:bandgap}, the system remains to be a
$\Qtilde$-insulator, meaning the \textit{bandgap}.  The CFL solution
ceases to exist at the point where these two curves meet, before which
a first-order phase transition to unpaired quark matter takes place as
indicated by the dotted vertical line (see discussions in
Ref.~\cite{Alford:2003fq} for details).

By substituting $\mue=0$, as is favored in the CFL phase with
electrons, for the above expressions, we readily realize that the
$bd$-$gs$ and $rs$-$bu$ dispersion relations are identical supposing
$\Delta_1=\Delta_2=\Delta$.  In fact, the $bd$-$gs$ and $rs$-$bu$
quark pairings are breached at the same time when
\begin{equation}
  |\deltamu{bd}{gs}|=|\deltamu{rs}{bu}|=\frac{\ms^2}{2\mu}>\Delta \,,
\end{equation}
which corresponds to the fact that in Fig.~\ref{fig:bandgap} the solid
and dashed lines cross at $\mue=0$.  This coincidence of the gapless
onset locations for $bd$-$gs$ and $rs$-$bu$ quarks is responsible for
instability for $A_{1,2}$ in the gCFL phase, as only $A_{1,2}$ can
excite a pair of $rs$ and $gs$ quarks at once (see
Fig.~\ref{fig:diag-rg-s}).  In other words, if the electron terms are
absent and $\mue$ is chosen to be positive, say $\sim10\MeV$, the
$A_{1,2}$ instability would disappear, while there would remain the
instability for other gluons.

\begin{figure}
\includegraphics[width=7cm]{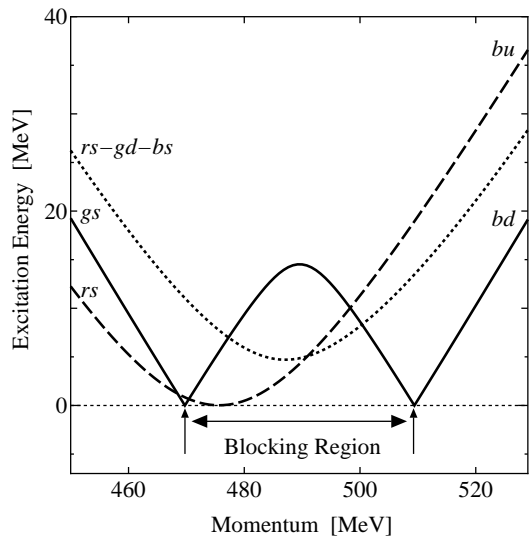}
\caption{Energy dispersion relations for $bd$-$gs$ (solid) and
$rs$-$bu$ (dashed) quarks at $\mssq=80\MeV$.  The dotted curve is one
that has the lowest energy among the 24 $ru$-$gd$-$bs$ dispersion
relations.}
\label{fig:dispersion}
\end{figure}

Once quark matter enters the gCFL phase with larger $\mssq$ than the
onset, the $bd$-$gs$ and $rs$-$bu$ quark dispersion relations are both
gapless but no longer degenerated.  The blocking momentum region for
$bd$-$gs$ quarks becomes wider with increasing $\mssq$ as shown by the
solid curve in Fig.~\ref{fig:dispersion} and $\Delta_1$ decreases
accordingly because quarks within the blocking region cannot
participate in pairing.  Since $rs$ and $bu$ quarks have nonzero
$\Qtilde$-charge, the blocking momentum region for $rs$-$bu$ quarks is
not allowed to be arbitrarily wider and is determined by the
requirement to cancel $\Qtilde$-charge brought by electrons whose
density is $\sim\mue^3$.  That means, the width of the blocking
momentum region for $rs$-$bu$ quarks is estimated as
$\delta p\sim \mue^3/3\bar{\mu}^2\ll\bar{\mu}$~\cite{Alford:2003fq}.
The actual form of the $rs$-$bu$ dispersion relation is, as a result,
kept to be almost quadratic with a tiny blocking region anywhere in
the gCFL phase as seen by the dashed curve in
Fig.~\ref{fig:dispersion}.

Although the $ru$-$gd$-$bs$ part is hard to anticipate \textit{a
priori}, there is no gapless mode involved in this sector.  The dotted
curve in Fig.~\ref{fig:dispersion} represents one of energy dispersion
relations for $ru$-$gd$-$bs$ quarks that has the lowest energy among
them.  It is clear from the figure that $ru$-$gd$-$bs$ quarks are all
gapped at $\mssq=80\MeV$ and it is also the case for larger $\mssq$.
Hence, the three (solid, dashed, and dot-dashed) lines drawn for
$bd$-$gs$, $rs$-$bu$, and $gu$-$rd$ quarks respectively in
Fig.~\ref{fig:bandgap} suffice for judging if gapless quarks are
present or not.


\subsection{Debye and Meissner screening masses}

In a medium at finite temperature and density the gauge fields are
screened by the polarization of charged thermal excitations.  The
screening effect on the gauge fields differs depending on the
longitudinal and transverse directions due to the lack of Lorentz
invariance.  The screening mass, which is the inverse of the screening
length, in the longitudinal direction is called the (chromo)electric
or Debye screening mass, and in (color) superconductors in particular
the transverse screening mass, the (chromo)magnetic or Meissner
screening mass in QED (QCD).  In the normal phase any perturbative
calculation leads to vanishing magnetic mass in the static limit.  In
the superconducting phase, on the other hand, a finite magnetic
screening mass results from the Higgs-Anderson mechanism and embodies
the Meissner effect so that superconductors can exclude the magnetic
field.

The Debye and Meissner masses are calculated from the eigenvalues of
the mass squared matrix in color defined by the self-energies;
\begin{align}
 m_{D,\alpha\beta}^2 &=
  -\lim_{q\to0}\Pi_{\alpha\beta}^{00}(\omega=0,\vec{q}), \\
 m_{M,\alpha\beta}^2 &= \half\lim_{q\to0}\bigl(\delta_{ij}-
  \widehat{q}_i \widehat{q}_j\bigr)\Pi_{\alpha\beta}^{ij}
  (\omega=0,\vec{q}),
\end{align}
where ($\alpha$,$\beta$) represent the color octet (or photon) indices
and we defined $\widehat{q}=\vec{q}/|\vec{q}|$, and
$\Pi_{\alpha\beta}^{\mu\nu}$ is the self-energy with the Lorentz
indices ($\mu$,$\nu$) for the gluons ($\alpha,\beta\le8$) or photon
($\gamma$ instead of $\alpha,\beta$) or their mixing.  (We always use
the subscript $\gamma$ for the sake of meaning photon.)  At the
one-loop level at high density, the quark loop contribution to
$\Pi_{\alpha\beta}^{\mu\nu}$, which is proportional to $\mu^2$, is
predominant, so we shall consider only the polarization tensor coming
from one-loop diagrams of quarks.

In the symmetric CFL state with $\ms=0$, the mass squared matrix is
diagonal in color and all eight gluons have the same screening mass.
The Debye and Meissner masses in a color superconductor have been
calculated diagrammatically in Ref.~\cite{Rischke:2000qz} and we shall
quote the results here;
\begin{align}
 m_D^2 &= \frac{21-8\ln 2}{6}m_g^2 \simeq 2.576 m_g^2 \,,
\label{eq:Debye}\\
 m_M^2 &= \frac{1}{3} m_D^2 \simeq 0.859 m_g^2
\label{eq:Meissner}
\end{align}
for all eight gluons.  Here $m_g^2= g^2\mu^2/6\pi^2$ and $g$ is the
strong coupling constant.  Note that our definition of $m_g^2$ is
different by a factor $N_f=3$ from the definition given in
Ref.~\cite{Rischke:2000qz}.

Once mixing between gluons and photon, which exists among $A_8$ and
$A_\gamma$ in the CFL phase at $\ms=0$, and generally among $A_3$,
$A_8$, and $A_\gamma$ for $\ms\neq0$, is taken into account, the
screening masses are to be modified.  They actually have been
evaluated analytically only when $\ms=0$~\cite{Schmitt:2003aa} and
after diagonalization two eigenvalues composed of $A_8$ and $A_\gamma$
read
\begin{align}
 \tilde{m}_{D,88}^2 &= \Bigl(1+\frac{4}{3}\frac{e^2}{g^2}\Bigr)m_D^2
  \,,\\
 \tilde{m}_{M,88}^2 &= \Bigl(1+\frac{4}{3}\frac{e^2}{g^2}\Bigr)m_M^2
  \,,
\label{eq:Meissner_8}\\
 \tilde{m}_{D,\gamma\gamma}^2 &= \tilde{m}_{M,\gamma\gamma}^2 = 0
  \,,
\label{eq:zero_photon}
\end{align}
where $e$ is the electromagnetic coupling constant.  The last relation
(\ref{eq:zero_photon}) follows from the facts that the condensates
(\ref{eq:diquark}) leave a ${\rm U}(1)_\Qtilde$ symmetry unbroken
(that is; $\tilde{m}_{M,\gamma\gamma}^2=0$), and that the system is
a $\Qtilde$-insulator in which $\Qtilde$-charged excitations are all
gapped (that is; $\tilde{m}_{D,\gamma\gamma}^2=0$).

For later convenience we shall carefully look into how these
analytical expressions (\ref{eq:Debye}) and (\ref{eq:Meissner}) are
structured from distinct contributions of particles and antiparticles.
When quarks are massless, the quark propagator can be specifically
separated into the particle and antiparticle parts (see
Eq.~(\ref{eq:propagator})).  At the quark one-loop level, the gluon or
photon self-energies consist of the particle-particle,
particle-antiparticle, and antiparticle-antiparticle excitations.  (In
the case of the normal phase the particle-particle excitations in the
vector channels would be to be articulated as the particle-hole
excitations.)  The last contribution from only the antiparticle
excitations is just negligible so that we shall always omit it in our
discussion.

In the $2\times2$ space of Nambu-Gor'kov doubling, the quark
propagator has not only the diagonal (normal) components but also the
off-diagonal (abnormal) components proportional to the gap $\Delta$
that connect two quasi-particles.

The Debye mass arises only from the particle-particle contribution.
It can be further divided into the Nambu-Gor'kov diagonal and
off-diagonal parts as follows;
\begin{equation}
 m_D^2 = [m_D^2]_{\text{diag}} + [m_D^2]_{\text{off}} \,,
\end{equation}
where
\begin{align}
 & [m_D^2]_{\text{diag}} = \frac{9}{2}m_g^2 \,,
\label{eq:Debye_diag}\\
 & [m_D^2]_{\text{off}} = -\biggl(1+\frac{4}{3}\ln 2\biggr)
  m_g^2 \,.
\end{align}
The sum of these two parts properly reproduces (\ref{eq:Debye}) as it
should.

In evaluating the Meissner mass, on the other hand, it is essentially
important to note that the particle-antiparticle excitation produces a
significant contribution comparable to the particle-particle one.  The
Meissner mass consists of the diagonal and off-diagonal parts,
\begin{equation}
 m_M^2 = [m_M^2]_{\text{diag}} + [m_M^2]_{\text{off}} \,,
\end{equation}
that is further split into the particle-particle (p-p) and
particle-antiparticle (p-a) parts,
\begin{equation}
 [m_M^2]_{\text{diag}} = [m_M^2]_{\text{diag(p-p)}}
  + 2[m_M^2]_{\text{diag(p-a)}}
\end{equation}
with
\begin{align}
 [m_M^2]_{\text{diag(p-p)}} &= -\frac{1}{3} [m_D^2]_{\text{diag}}\,,
\label{eq:diag(p-h)}\\
 [m_M^2]_{\text{diag(p-a)}} &= \frac{3}{2}m_g^2
\label{eq:diag(p-a)}\,,
\end{align}
and
\begin{equation}
 [m_M^2]_{\text{off}} = [m_M^2]_{\text{off(p-p)}}
  =\frac{1}{3} [m_D^2]_{\text{off}}
\label{eq:off} \,,
\end{equation}
from which we can easily make sure $m_M^2 = \third m_D^2$ as is
well-known in the CFL phase~\cite{Rischke:2000qz,Son:1999cm}.  The
particle-antiparticle contribution in the off-diagonal part
$[m_M^2]_{\text{off(p-a)}}$ is vanishingly small, so we dropped it off
from the above relations.

It is of great importance to realize that the diagonal p-p
contribution to the Meissner mass is \textit{negative} one third of
the Debye mass counterpart, while the p-a excitations provide a
\textit{positive} contribution that is twice larger than the
p-p's. Adding them together we finally acquire the Meissner mass
contribution that is \textit{positive} one third of the Debye mass
contribution. In the off-diagonal part, in contrast, the situation is
much simpler and the ratio is positive one third itself.

In other words, in the diagonal part, the p-p loops always tend to
induce \textit{paramagnetism}, while \textit{diamagnetism} originates
from the p-a loops.  Usually in the superconducting phase, the
diamagnetic tendency is greater enough to bring about the Meissner
effect.  In gapless superconductors, however, antiparticles are never
gapless and only the p-p loops are abnormally enhanced due to gapless
quarks and their large density of states near the Fermi surface causes
the opposite phenomenon to the Meissner effect, i.e., chromomagnetic
instability.  Also, this can be seen from the relations
(\ref{eq:diag(p-h)}) and (\ref{eq:off}) which hold in the presence of
finite $\ms$ or even in the gCFL phase.  In the gCFL phase, a large
density of states leads to a large Debye mass.  Then, through
(\ref{eq:diag(p-h)}) and (\ref{eq:off}), it should be accompanied by a
large \textit{negative} contribution to the Meissner mass squared,
which eventually results in chromomagnetic instability.

As for (\ref{eq:diag(p-a)}), because the $\ms$ dependence of
antiparticle excitations is suppressed by $\ms/\mu$, the relation is
hardly changed for any $\ms\ll\mu$.


\section{computation of self-energies}
\label{sec:self-energies}

In this section we shall diagrammatically compute the calculable parts
of one-loop self-energies for gluons and photon and derive the
analytical expression for the singular part of the Debye and Meissner
masses.

Using the mean-field quark propagator $S(p)$, and the vertex matrices,
$\Gamma^\mu_\alpha$, in color, flavor, spin, and Nambu-Gor'kov space,
we can write the one-loop self-energies in a general form as
\begin{equation}
 -i\Pi_{\alpha\beta}^{\mu\nu}(q)=-\frac{1}{2}\int^T\!\!\!
  \frac{d^4p}{(2\pi)^4}\,\tr\, \Gamma^\mu_\alpha S(q+p)
  \Gamma^\nu_\beta S(p) \,.
\label{eq:self-energy}
\end{equation}
The vertices $\Gamma^\mu_\alpha$ take a matrix form,
\begin{equation}
 \Gamma^\mu_\alpha =\left[ \begin{array}{cc}
  -ig\gamma^\mu T_\alpha \delta_{ij} & 0 \\
  0 & ig\gamma^\mu T_\alpha \delta_{ij}
 \end{array} \right]
\end{equation}
for gluons ($\alpha=1,\dots,8$) where $T_\alpha$ are the
${\rm SU}(3)_{\text{color}}$ generators defined by the Gell-Mann
matrices in color space and normalized as
$\tr T_\alpha T_\beta=\half\delta_{\alpha\beta}$.  For photon the
vertex reads
\begin{equation}
 \Gamma^\mu_\gamma =\left[ \begin{array}{cc}
  -ie\gamma^\mu Q_{ij} & 0 \\
  0 & ie\gamma^\mu Q_{ij}
 \end{array} \right],
\end{equation}
which is unity in color.  We see that the flavor is not changed at any
vertices but the color can be converted through the off-diagonal
components in $T_1$, $T_2$, $T_4$, $T_5$, $T_6$, and $T_7$.  We call
the gluons corresponding to $T_3$ and $T_8$ as the color-diagonal
gluons.

The quark propagator is divided into the particle and antiparticle
parts by the energy projection operators
$\Lambda_p^\pm=\half(1\pm\gamma^0\vec{\gamma}\cdot\widehat{p})$.
The explicit propagator in the $bd$-$gs$ sector, for instance, can be
written down after some rearrangement of the Nambu-Gor'kov (1,2)
(where 2 is assigned to the Nambu-Gor'kov doubler in our convention)
and color-flavor ($bd$,$gs$) indices, that is, in the basis
($bd$-1,$gs$-2,$gs$-1,$bd$-2), we have
\begin{equation}
\begin{split}
 &-iS(p) = \left[ \begin{array}{cccc}
  \Lambda^-_p & 0 & 0 & 0 \\
  0 & \Lambda^+_p & 0 & 0 \\
  0 & 0 & \Lambda^-_p & 0 \\
  0 & 0 & 0 & \Lambda^+_p
  \end{array} \right]\gamma^0
    \left[ \begin{array}{cc}
     S^{\rm a}_{\text{$bd$-$gs$}}(p) & 0\\
     0 & S^{\rm a}_{\text{$gs$-$bd$}}(p)
    \end{array} \right] \\
 &\qquad +\left[ \begin{array}{cccc}
  \Lambda^+_p & 0 & 0 & 0 \\
  0 & \Lambda^-_p & 0 & 0 \\
  0 & 0 & \Lambda^+_p & 0 \\
  0 & 0 & 0 & \Lambda^-_p
  \end{array} \right]\gamma^0
    \left[ \begin{array}{cc}
     S^{\rm p}_{\text{$bd$-$gs$}}(p) & 0\\
     0 & S^{\rm p}_{\text{$gs$-$bd$}}(p)
    \end{array} \right] \,,
\end{split}
\label{eq:propagator}
\end{equation}
where the antiparticle part is
\begin{equation}
\begin{split}
 S^{\rm a}_{\text{$A$-$B$}}(p) &= \frac{1}{\bigl(p_0+\aenergy{A}{B}
  \bigr)\bigl(p_0-\aenergy{B}{A}\bigr)} \\
&\qquad\times
\left[ \begin{array}{cc}
  p_0-(p+\mu_B)  &  -i\Delta_1\gamma_5\gamma^0 \\
  -i\Delta_1\gamma_5\gamma^0  &  p_0+(p+\mu_A)
  \end{array} \right] \,,
\end{split}
\label{eq:propagator_a}
\end{equation}
and the particle part is
\begin{equation}
\begin{split}
 S^{\rm p}_{\text{$A$-$B$}}(p) &= \frac{1}{\bigl(p_0+\energy{A}{B}
  \bigr)\bigl(p_0-\energy{B}{A}\bigr)} \\
&\qquad\times
\left[ \begin{array}{cc}
  p_0+(p-\mu_B)  &  -i\Delta_1\gamma_5\gamma^0 \\
  -i\Delta_1\gamma_5\gamma^0  &  p_0-(p-\mu_A)
  \end{array} \right] \,.
\end{split}
\label{eq:propagator_p}
\end{equation}

In this representation the structure of the Dirac indices becomes much
simpler and is given by $\Lambda^\pm_p\gamma^0$ and $\gamma_5\gamma^0$
attached with $\Delta$.  Then, it is important to notice that the
$(\mu,\nu)$ dependence in the integrand of (\ref{eq:self-energy})
comes from the trace over the Dirac indices alone, that is
\begin{align}
 {\cal T}_{\text{diag(p-p)}}^{\mu\nu}(p;q) &=
  \tr\bigl[\gamma^\mu\Lambda_{q+p}^+\gamma^0\gamma^\nu\Lambda_p^+
  \gamma^0\bigr],
\label{eq:trace_diag_ph}\\
 {\cal T}_{\text{diag(p-a)}}^{\mu\nu}(p;q) &=
  \tr\bigl[\gamma^\mu\Lambda_{q+p}^+\gamma^0\gamma^\nu\Lambda_p^-
  \gamma^0\bigr]
\label{eq:trace_diag_pa}
\end{align}
for p-p and p-a loops made of the Nambu-Gor'kov diagonal components of
the propagator (\ref{eq:propagator_a}) and (\ref{eq:propagator_p}).
For the Nambu-Gor'kov off-diagonal components we have likewise;
\begin{align}
 {\cal T}_{\text{off(p-p)}}^{\mu\nu}(p;q) &=
  \tr\bigl[\gamma^\mu\Lambda_{q+p}^+\gamma_5
  \gamma^\nu\Lambda_p^-\gamma_5\bigl],
\label{eq:trace_off_ph}\\
 {\cal T}_{\text{off(p-a)}}^{\mu\nu}(p;q) &=
  \tr\bigl[\gamma^\mu\Lambda_{q+p}^+\gamma_5
  \gamma^\nu\Lambda_p^+\gamma_5\bigl] \,.
\label{eq:trace_off_pa}
\end{align}


\subsection{Singularities in $A_{1,2}$}

\begin{figure}
\includegraphics[width=7cm]{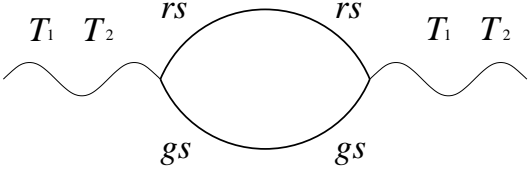}
\caption{An example of diagrams consisting of the diagonal components
contributing to the self-energy for $A_{1,2}$.}
\label{fig:diag-rg-s}
\end{figure}

\begin{figure}
\includegraphics[width=7cm]{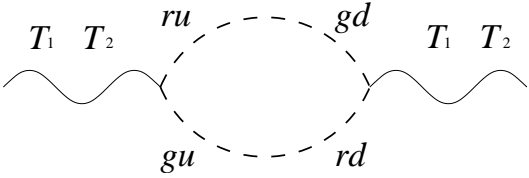}
\caption{An example of diagrams consisting of the off-diagonal
components contributing to the self-energy for $A_{1,2}$.  Neither
$ru$-$gd$ nor $gu$-$rd$ propagation involves gapless quarks, so that
this diagram could not lead to any singular behavior in the gCFL
phase.}
\label{fig:diag-rg-ud}
\end{figure}

We will first discuss $A_1$ and $A_2$ that are degenerated.  The
generators in color, $T_1$ and $T_2$, have a non-vanishing component
only between red and green and the flavor is kept unchanged at the
vertices, so that the self-energy for $A_{1,2}$ stems from four
diagrams;
\begin{equation}
\begin{split}
 \Pi_{11}^{\mu\nu} &=
  \Pidiag{\mu\nu}{rg;s}+\Pidiag{\mu\nu}{rg;u} \\
 &\qquad\quad +\Pidiag{\mu\nu}{rg;d}
  +\Pioff{\mu\nu}{ru}{gd}{gu}{rd}.
\end{split}
\end{equation}
Here, diag($rg;s$) means the quark loop composed of the diagonal
component of $rs$ and $gs$ quarks (see Fig.~\ref{fig:diag-rg-s}) and
diag($rg;u$) and diag($rg;d$) should be understood in the same manner.
The last contribution, off($ru$-$gd$;$gu$-$rd$), represents the
self-energy contribution from the off-diagonal components of $ru$-$gd$
and $gu$-$rd$ propagation, that is, the loop made up with $ru$ quarks
turning into $gd$ and $rd$ quarks turning into $gu$ as shown in
Fig.~\ref{fig:diag-rg-ud}.

\begin{figure}
\includegraphics[width=7cm]{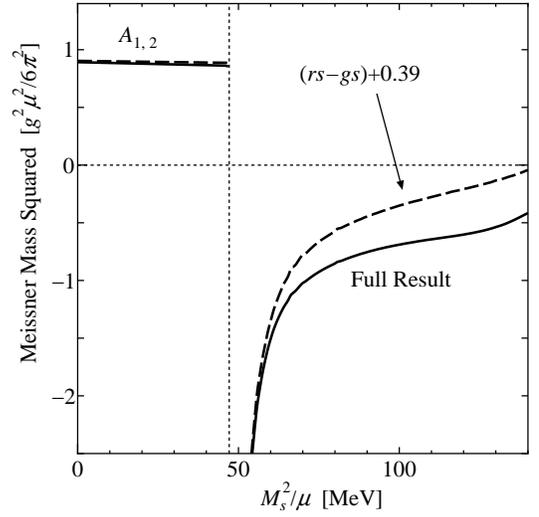}
\caption{Comparison between the $rs$-$gs$ contribution with an offset
$0.39$ and the full result for the Meissner mass squared for
$A_{1,2}$.  This means that the gross feature near the gCFL onset is
provided by the $rs$-$gs$ diagram alone.}
\label{fig:rs_gs-full}
\end{figure}

In the case of $A_1$ (and $A_2$ equivalently) only
$\Pidiag{\mu\nu}{rg;s}$ may contain gapless quarks in the gCFL phase
in both lines of the loop diagram, and is expected to provide a
singular contribution.  As a matter of fact, the singular part
presumably dominates over the screening mass behavior near the gapless
onset as a function of $\ms$.  This expectation is confirmed by the
comparison between functional forms of only the $rs$-$gs$ polarization
and the full result derived from the potential curvature we are
dealing with in the next section.  From Fig.~\ref{fig:rs_gs-full},
taking in the full results in advance, we can clearly observe that the
physics near the gCFL onset is to be described by the $rs$-$gs$
excitation only.

The Debye screening mass, or $\Pi^{00}$, has only the
particle-particle part (and the antiparticle-antiparticle part that is
negligible) because ${\cal T}_{\text{diag(p-a)}}^{00}(p;0)
={\cal T}_{\text{off(p-a)}}^{00}(p;0)=0$  and is written as
\begin{equation}
\begin{split}
 &\Pidiag{00}{rg;s}(q\to0) \\ &=
  -\frac{g^2}{8\pi^2}\int dp\, p^2\,
  {\cal T}_{\text{diag(p-p)}}^{00}(p;0)\,
  {\cal U}_{\text{diag($rg;s$)}}(p) \,,
\end{split}
\end{equation}
where ${\cal T}_{\text{diag(p-h)}}^{00}(p;0)=2$ and we defined
\begin{equation}
\begin{split}
 & {\cal U}_{\text{diag($rg;s$)}} =
  \fph[\energy{bu}{rs};bd,gs,bu] \\
 & +\fph[-\energy{rs}{bu};bd,gs,bu]
   +\fph[\energy{bd}{gs};bu,rs,bd] \\
 & +\fph[-\energy{gs}{bd};bu,rs,bd].
\label{eq:rg-s-00-f}
\end{split}
\end{equation}
The p-p integrand $\fph$ is explicitly given by
\begin{equation}
 \begin{split}
 & \fph[\energy{A}{B};C,D,E] \\
 & =\frac{\bigl[\energy{A}{B} + (p\!-\!\mu_E)\bigr]
        \bigl[\energy{A}{B} + (p\!-\!\mu_C)\bigr]}
  {\bigl[\energy{A}{B}\!+\!\energy{B}{A}\bigr]
        \bigl[\energy{A}{B} \!-\!\energy{C}{D}\bigr]
        \bigl[\energy{A}{B} \!+\!\energy{D}{C}\bigr]}
 \tanh\biggl[\frac{\energy{A}{B}}{2T}\biggr]
 \end{split}
\label{eq:fph}
\end{equation}
with the energy dispersion relation $\energy{A}{B}$ defined in
(\ref{eq:dispersion}).

In the CFL phase, as we have already mentioned, the $bd$-$gs$ and
$rs$-$bu$ quark dispersion relations are identical and there are
vanishing combinations in the denominator of (\ref{eq:fph}) like
$\energy{gs}{bd}(q+p)-\energy{rs}{bu}(p)\to0$ as $q\to0$.  The
numerator goes to zero in the same limit of $q\to0$ and consequently
it amounts to the derivative with respect to $\varepsilon$ acting onto
the distribution function, i.e.,
$(\partial/\partial\energy{A}{B})\tanh[\energy{A}{B}/2T]$.  At zero
temperature this is proportional to the delta function
$\delta(\energy{A}{B})$ that takes a finite value only when gapless
quarks (that is; $\energy{A}{B}(p)=0$ for some $p$) are present.

Using the solution (\ref{eq:CFL_mu3}) and (\ref{eq:CFL_mu8}) and
assuming $\Delta_1=\Delta_2=\Delta$ that is known to be a good
approximation in the CFL phase, the polarization at zero temperature
can simplify as
\begin{equation}
 \begin{split}
 & \Pidiag{00}{rg;s} \\
 &=-\frac{g^2}{4\pi^2}\int dp\, p^2 \biggl\{
  \frac{[\varepsilon - (p-\bar{\mu})]^2}{2\varepsilon^2}\delta\Bigl(
  \varepsilon-{\txt\frac{\ms^2}{2\mu}}\Bigr)
  +\frac{\Delta^2}{2\varepsilon^3} \biggr\}
 \end{split}
\label{eq:rg-s-00}
\end{equation}
with $\bar{\mu}=\mu-\ms^2/6\mu$ and
$\varepsilon=\sqrt{(p-\bar{\mu})^2+\Delta^2}$.  This expression is
valid only up to the onset where the gCFL starts.  Actually, right at
the onset where $\ms^2/2\mu=\Delta$, the energy dispersion relation
for $gs$-$bd$ and $rs$-$bu$ quarks is quadratic near the Fermi
momentum and approximated as $\varepsilon(p)-\ms^2/2\mu\simeq
(p-\bar{\mu})^2/2\Delta$.  Then, the $p$-integration for the first
term of (\ref{eq:rg-s-00}) picks up the density of states at
$p=\bar{\mu}$ and ends up with an infrared divergence.  We would
emphasize that this singular behavior originates from the electron
pressure which ensures $\mue=0$.  Otherwise, two quarks running
through the loop diagram would not be gapless simultaneously at the
gCFL onset.

In the CFL region before the gCFL occurs, it is rather the second term
of (\ref{eq:rg-s-00}) which is a major contribution, which can be
further approximated as
\begin{equation}
 \Pidiag{00}{rg;s}=-\frac{g^2\mu^2}{4\pi^2}=-\frac{3}{2}m_g^2
\end{equation}
with any logarithmic divergence neglected.  The logarithmic terms are,
in fact, given by $\sim\Delta^2\ln[\Lambda/\Delta]$ that is much
smaller than the leading terms of order $\bar{\mu}^2$ and corrections
are $\sim$1\% for our choice of $\Lambda=800\MeV$ and
$\Delta_0=25\MeV$.  We will always drop any logarithmic terms
$\sim\ln[\Lambda/\Delta]$ throughout this paper.

Hence, the contribution to the Debye mass in the CFL side is obtained
as
\begin{equation}
 [m_D^2]_{\text{diag($rg;s$)}} = \frac{3}{2}m_g^2 \,.
\end{equation}
Interestingly enough, this result exactly corresponds to one-flavor
contribution (i.e.\ one-third) of (\ref{eq:Debye_diag}) and does not
have $\ms$-dependence until the gCFL onset.

We can immediately extend our discussion to the Meissner mass once if
we analyze the ($\mu$,$\nu$) structure in the polarization.  The p-p
contribution to the Meissner mass is
\begin{equation}
 \begin{split}
 & \frac{1}{2}\bigl(\delta_{ij}-\widehat{q}_i \widehat{q}_j
  \bigr) {\cal T}_{\text{diag(p-p)}}^{ij}(p;0) \\
 =& \frac{1}{2}\bigl(\delta_{ij}-\widehat{q}_i \widehat{q}_j
  \bigr) \times 2\widehat{p}^i\widehat{p}^j
 = 1-(\widehat{q}\cdot\widehat{p})^2 \\
 \to& \frac{2}{3}
 = \frac{1}{3} {\cal T}_{\text{diag(p-p)}}^{00}(p;0).
 \end{split}
\label{eq:calc_meissner}
\end{equation}
In the third line we made use of averaging over the angle integration
that leads to one third, that is, $\int\!d^3p\,(\widehat{q}\cdot
\widehat{p})^2\,{\cal U}(p)=\third\int\!d^3p\,{\cal U}(p)$.  In
evaluating the Debye and Meissner masses ${\cal
U}_{\text{diag($rg;s$)}}(p)$ is common, so that from the above
relation we reach the general relation,
\begin{equation}
 [m_M^2]^{\text{(p-p)}}_{\text{diag($rg;s$)}} =
  -\frac{1}{3}[m_D^2]_{\text{diag($rg;s$)}}, 
\label{eq:m_M_diag}
\end{equation}
that is completely consistent with (\ref{eq:diag(p-h)}).

The p-a part generates ultraviolet divergent terms proportional to
$\Lambda^2$, $\Delta^2\ln[\Lambda/\Delta]$, and
$(\mssq)^2\ln[\Lambda/\Delta]$~\cite{Alford}.  In the present work, as
in the treatment of the Debye mass, we ignore any logarithmic
divergences.  We can rewrite $\fph$ into $\fpa$ immediately, in view
of the difference between (\ref{eq:propagator_a}) and
(\ref{eq:propagator_p}), by replacing corresponding $\energy{A}{B}$ by
$\aenergy{A}{B}$, and $p\!-\!\mu_A$ by $-(p\!+\!\mu_A)$.  A similar
analysis to (\ref{eq:calc_meissner}) on the Dirac trace part leads to
a factor as follows;
\begin{equation}
 \begin{split}
 & \frac{1}{2}\bigl(\delta_{ij}-\widehat{q}_i \widehat{q}_j
  \bigr) {\cal T}_{\text{diag(p-a)}}^{ij}(p;0) \\
 =& \frac{1}{2}\bigl(\delta_{ij}-\widehat{q}_i \widehat{q}_j
  \bigr) \times 2\bigl[\delta^{ij}-\widehat{p}^i\widehat{p}^j\bigr]
 = 1+(\widehat{q}\cdot\widehat{p})^2
  \to \frac{4}{3}.
 \end{split}
\label{eq:calc_meissner2}
\end{equation}
Then, after some calculations along the same line as
Refs.~\cite{Huang:2004bg,Rischke:2000qz} with the approximation that
$\Delta$ and $\ms$ are all neglected, we acquire
\begin{equation}
 \begin{split}
 & [m_M^2]^{\text{(p-a)}}_{\text{diag($rg;s$)}} \\
 = & \frac{1}{2}\lim_{q\to0}\bigl(\delta_{ij}-\widehat{q}_i
  \widehat{q}_j \bigr) \Bigr[\Pidiag{ij}{rg;s}
  \Bigr]^{\text{(p-a)}} \\
 \simeq & \frac{1}{2}m_g^2 - \frac{\Lambda^2}{12\pi^2} \,,
 \end{split}
\label{eq:meissner_p-a}
\end{equation}
that is consistent with (\ref{eq:diag(p-a)}) divided by the flavor
factor $N_{\rm f}=3$ up to the ultraviolet divergence.  

After all, the sum over the p-p, p-a, and a-p parts finally amounts to
\begin{equation}
\begin{split}
 & [m_M^2]_{\text{diag($rg;s$)}} =
  [m_M^2]^{\text{(p-p)}}_{\text{diag($rg;s$)}} \\
 & \qquad +2\times [m_M^2]^{\text{(p-a)}}_{\text{diag($rg;s$)}}
  + \text{(subtraction)} = \frac{1}{2}m_g^2
\end{split}
\end{equation}
in the CFL phase, where ``subtraction'' is a term added by hand to
remove the $\Lambda^2$-term which should be renormalized.
Eq.~(\ref{eq:meissner_p-a}) indicates
\begin{equation}
 \text{(subtraction)} = \frac{\Lambda^2}{6\pi^2}
\label{eq:subtraction}
\end{equation}
per one polarization diagram of Nambu-Gor'kov diagonal
particle-antiparticle excitations that should contain ultraviolet
divergences.


\subsection{No singularities in $A_{4,5}$ and $A_{6,7}$}

The analyses in the previous subsection give us information about the
origin of singularities around the gCFL onset.  When two quarks
involved in the loop diagram becomes gapless at the same time, the
divergent density of states compels us to face the imaginary Meissner
mass.  We can easily comprehend that this kind of singularity near the
gapless onset is absent for $A_{4,5}$ and $A_{6,7}$.

The self-energies are diagrammatically decomposed as
\begin{equation}
\begin{split}
 \Pi_{44}^{\mu\nu} &=
  \Pidiag{\mu\nu}{br;s}+\Pidiag{\mu\nu}{br;u} \\
 &\qquad\quad +\Pidiag{\mu\nu}{br;d}
  +\Pioff{\mu\nu}{bu}{rs}{ru}{bs}
\end{split}
\end{equation}
and
\begin{equation}
\begin{split}
 \Pi_{66}^{\mu\nu} &=
  \Pidiag{\mu\nu}{gb;s}+\Pidiag{\mu\nu}{gb;u} \\
 &\qquad\quad +\Pidiag{\mu\nu}{gb;d}
  +\Pioff{\mu\nu}{gd}{bs}{bd}{gs}.
\end{split}
\end{equation}
None of these above polarizations is composed of would-be gapless
quarks only, i.e.\ $bd$-$gs$ and $rs$-$bu$ quarks only.  Therefore, we
can conclude at this stage of analyses that $A_{4,5}$ and $A_{6,7}$ do
not exhibit chromomagnetic instability at least in the vicinity of the
gCFL onset.


\subsection{Singularities in $A_{3,8,\gamma}$}

\begin{figure}
\includegraphics[width=7cm]{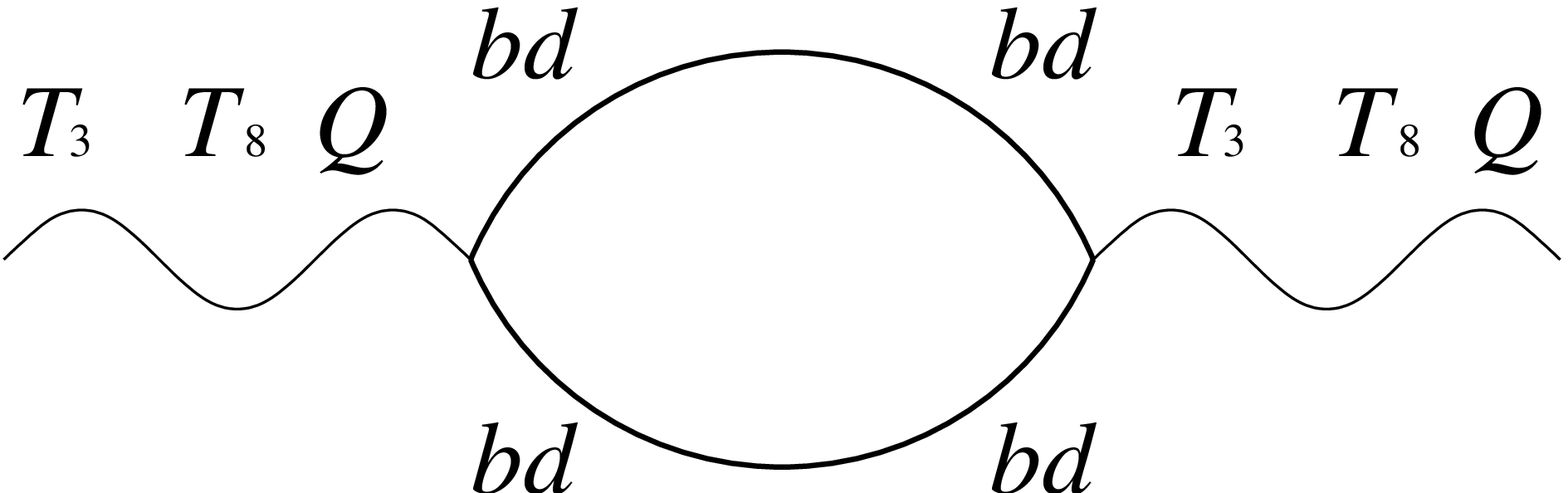}
\caption{An example of diagrams consisting of the diagonal components
contributing to the self-energy for diagonal gluons and photon.}
\label{fig:diag-bd}
\end{figure}

\begin{figure}
\includegraphics[width=7cm]{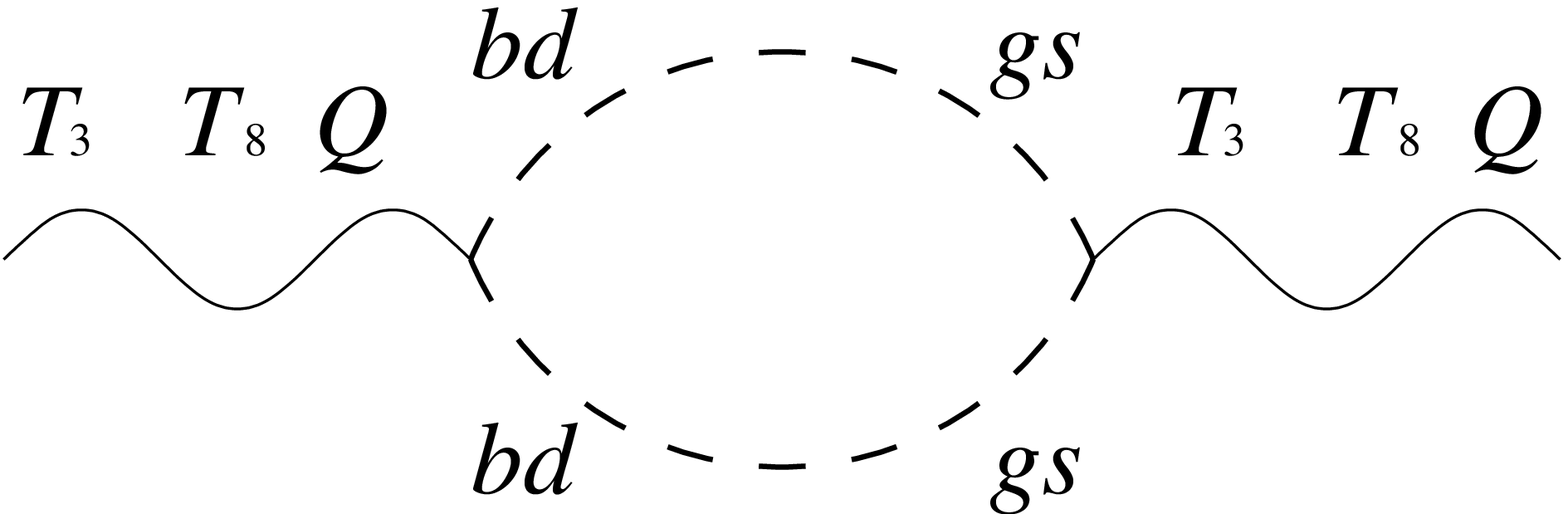}
\caption{An example of diagrams consisting of the off-diagonal
components contributing to the self-energy for diagonal gluons and
photon.}
\label{fig:diag-bd-gs}
\end{figure}

For color-diagonal gluons $A_3$, $A_8$, and photon $A_\gamma$, neither
color nor flavor is changed at the vertices.  This means that 9
diagonal and 6 off-diagonal diagrams are possible.  For example, the
self-energy for $A_8$, for instance, has two types of singularities;
one is associated with the $bd$-$gs$ gapless quark dispersion
relation,
\begin{equation}
\begin{split}
 \bigl[\Pi_{88}^{\mu\nu}\bigr]_{\text{$bd$-$gs$}} &=
  \frac{1}{3} \Pidiag{\mu\nu}{gg;s} + \frac{4}{3}\Pidiag{\mu\nu}{bb;d}\\
 &\qquad - \frac{2}{3} \Pioff{\mu\nu}{bd}{gs}{bd}{gs} \,,
\end{split}
\label{eq:sing1}
\end{equation}
and the other is associated with the $rs$-$bu$ gapless quark (almost
quadratic) dispersion relation,
\begin{equation}
\begin{split}
 \bigl[\Pi_{88}^{\mu\nu}\bigr]_{\text{$rs$-$bu$}} &=
  \frac{1}{3} \Pidiag{\mu\nu}{rr;s} + \frac{4}{3}\Pidiag{\mu\nu}{bb;u}\\
 &\qquad - \frac{2}{3} \Pioff{\mu\nu}{rs}{bu}{rs}{bu} \,.
\end{split}
\label{eq:sing2}
\end{equation}
The one-loop diagrams corresponding to $\Pidiag{\mu\nu}{bb;d}$ and
$\Pioff{\mu\nu}{bd}{gs}{bd}{gs}$ are shown in Figs.~\ref{fig:diag-bd}
and \ref{fig:diag-bd-gs} to take some examples.

For completeness we shall enumerate the rest here.  The non-singular
contributions to the $A_8$ polarization are
\begin{equation}
\begin{split}
 & \bigl[\Pi_{88}^{\mu\nu}\bigr]_{\text{non-singular}}
 =\frac{1}{3}\Pidiag{\mu\nu}{rr;u}
   + \frac{1}{3}\Pidiag{\mu\nu}{rr;d} \\
 & + \frac{1}{3}\Pidiag{\mu\nu}{gg;u}
   + \frac{1}{3}\Pidiag{\mu\nu}{gg;d}
   + \frac{4}{3}\Pidiag{\mu\nu}{bb;s} \\
 & + \frac{1}{3}\Pioff{\mu\nu}{gu}{rd}{gu}{rd}
   + \frac{1}{3}\Pioff{\mu\nu}{ru}{gd}{ru}{gd} \\
 & - \frac{2}{3}\Pioff{\mu\nu}{gd}{bs}{gd}{bs}
   - \frac{2}{3}\Pioff{\mu\nu}{bs}{ru}{bs}{ru} \,,
\label{eq:non-singular}
\end{split}
\end{equation}

The diagonal contributions, as shown in Fig.~\ref{fig:diag-bd}, are
easily available from the previous results only by replacing the
color-flavor indices appropriately and adding $\half$ to adjust the
combinatorial factor.  It should be noted that instability occurs in
this case whenever any of quarks becomes gapless because two virtually
excited quarks are identical in the color-diagonal channels.

The off-diagonal expression is also deduced simply by means of proper
arrangement of the color-flavor indices and replacement of the
numerator of (\ref{eq:fph})
as $\energy{A}{B}\pm(p-\mu_B)\to-i\Delta_{AB}$.
Because $\energy{A}{B}\pm(p-\mu_B)\simeq\Delta_{AB}$ in the vicinity
of the Fermi surface $p\!\sim\!\mu$ that is dominating over the
$p$-integration, the functional form of the off-diagonal contribution
is essentially the same as the diagonal contributions.  There are two
important features to remark here, though.

First, the self-energy contributions coming from the off-diagonal loop
are free from ultraviolet quadratic divergence even in p-a
excitations.  It is because the off-diagonal propagator is
proportional to, not $\energy{A}{B}\pm(p-\mu_B)$, but $\Delta_{AB}$
that is not rising with increasing $p$.  Thus, the subtraction term
proportional to $\Lambda^2$ is no longer necessary.

Second, from the trace over the Dirac indices (\ref{eq:trace_off_ph})
and (\ref{eq:trace_off_pa}) we notice that
\begin{align}
 {\cal T}_{\text{off}}^{00} &= -{\cal T}_{\text{diag}}^{00},
\label{eq:relation_diag_off00}\\
 {\cal T}_{\text{off}}^{ij} &= {\cal T}_{\text{diag}}^{ij}
\label{eq:relation_diag_offij}
\end{align}
for both p-p and p-a cases, that means the diagonal and off-diagonal
diagrams contribute to the Debye and Meissner masses differently.
Using (\ref{eq:relation_diag_off00}), (\ref{eq:relation_diag_offij})
and from a similar analysis to (\ref{eq:calc_meissner}), we
immediately see the relation,
\begin{equation}
 [m_M^2]_{\text{off($bd$-$gs$;$bd$-$gs$)}}
  =\frac{1}{3}[m_D^2]_{\text{off($bd$-$gs$;$bd$-$gs$)}},
\label{eq:m_M_off}
\end{equation}
that is consistent with (\ref{eq:off}).  The sign difference between
(\ref{eq:m_M_diag}) and (\ref{eq:m_M_off}) originates from the
negative sign in (\ref{eq:relation_diag_off00}).  The p-a off-diagonal
contribution to the Meissner mass squared turns out to be negligible.

In addition, the negative sign difference between
(\ref{eq:relation_diag_off00}) and (\ref{eq:relation_diag_offij}) can
explain the results for the ``neutral'' and ``charged'' condensates
discussed in a two-species model in Ref.~\cite{Alford:2005qw}.  As we
have mentioned above, the functional form of $\Pidiag{\mu\nu}{gg;s}$,
$\Pidiag{\mu\nu}{bb;d}$, and $\Pioff{\mu\nu}{bd}{gs}{bd}{gs}$ are
almost identical up to the combinatorial factor $\half$.  Let's
consider the ``neutral'' case fictitiously following the argument in
Ref.~\cite{Alford:2005qw}, for which we \textit{assume} that $bd$ and
$gs$ quarks had the opposite charge corresponding to the two-species
model composed of $bd$ and $gs$.  Then, the polarization related to
the screening masses would be
$\Pi^{\mu\nu}=\Pidiag{\mu\nu}{gg;s}+\Pidiag{\mu\nu}{bb;d}-\Pioff{\mu\nu}{bd}{gs}{bd}{gs}$.
From (\ref{eq:relation_diag_off00}) we see that the Debye mass
positively diverges at the gapless onset, while the singularities in
the Meissner mass cancel to vanish.  In the ``charged'' case, on the
other hand, we \textit{assume} that $bd$ and $gs$ quarks had the same
charge.  In the same way, the polarization leading to the screening
masses would be
$\Pi^{\mu\nu}=\Pidiag{\mu\nu}{gg;s}+\Pidiag{\mu\nu}{bb;d}+\Pioff{\mu\nu}{bd}{gs}{bd}{gs}$.
Obviously, the Debye mass has no singularity at the gapless onset, but
the Meissner mass negatively diverges.  This is exactly what was found
in Ref.~\cite{Alford:2005qw}.  In QCD, however, the diquark condensate
is actually a mixture of the ``neutral'' and ``charged'' condensates
and the Debye and Meissner masses both diverge positively and
negatively from the common origin at the gCFL onset.

As a final remark of this section, we shall draw attention to the
contrasting difference between the singular parts (\ref{eq:sing1})
related to $bd$-$gs$ quarks and (\ref{eq:sing2}) related to $rs$-$bu$
quarks.  In general, the Meissner mass negatively diverges on the
$bd$-$gs$, $rs$-$bu$, and $gu$-$rd$ boundary lines drawn in
Fig.~\ref{fig:bandgap}.  When $\ms$ grows larger, as seen in
Fig.~\ref{fig:dispersion}, the blocking region of the $bd$-$gs$
dispersion relation becomes wider and the singular character in
(\ref{eq:sing1}) weakens.  In contrast, $\mue$ is kept to be very
close to the $rs$-$bu$ line in Fig.~\ref{fig:bandgap} for all $\ms$
and in all the gCFL region (\ref{eq:sing2}) remains to provide a
negatively huge contribution to the Meissner mass.  If it were not for
electrons, $\mue$ would be within the bandgap and (\ref{eq:sing2})
would not be singular at all.  In this sense, the singularity of
(\ref{eq:sing2}) is induced by the presence of the electron terms
which force $\mue$ to stay in the singular region slightly below the
dashed line in Fig.~\ref{fig:bandgap}.  To evaluate that huge
contribution correctly from (\ref{eq:sing2}), it is essential to
resolve a tiny blocking region $\delta p\sim\mue^3/3\bar{\mu}^2$ with
a great accuracy, which is next to impossible technically.

\begin{figure}
\includegraphics[width=7cm]{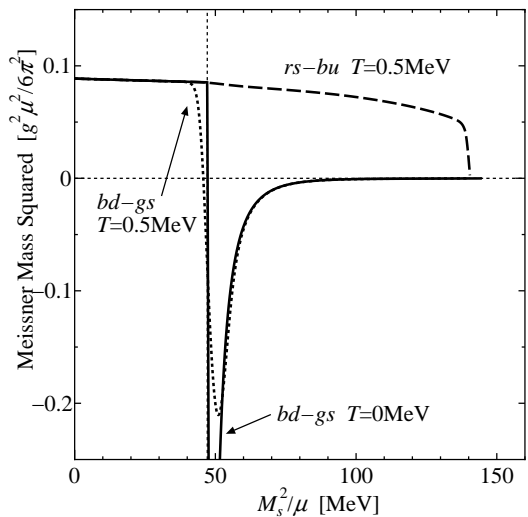}
\caption{Each contribution from possible singular diagrams at $T=0$
and $0.5\MeV$.  In the figure $bd$-$gs$ and $rs$-$bu$ represent
contributions to the Meissner mass from
$[\Pi^{\mu\nu}_{88}]_{\text{$bd$-$gs$}}$ and
$[\Pi^{\mu\nu}_{88}]_{\text{$rs$-$bu$}}$.}
\label{fig:sing}
\end{figure}

Although the strength of singularities (\ref{eq:sing1}) and
(\ref{eq:sing2}) at the gCFL onset is exactly equal at $T=0$, the
temperature dependence is drastically different;  as seen from
Fig.~\ref{fig:sing}, (\ref{eq:sing2}) is no longer singular even at
$T=0.5\MeV$.  This is because at finite temperature $rs$-$bu$ quarks
not from the blocking region but from thermal excitations can cancel
the $\Qtilde$-charge driven by electrons.  There needs not to be a
blocking region associated with $rs$-$bu$ quarks therefore.  Actually,
the $rs$-$bu$ quark density is estimated as
$\sqrt{2\pi\Delta T}\bar{\mu}^2/\pi^2$ when its dispersion relation is
quadratic, and it can balance out the electron density
$\mue^3/3\pi^2$ even at $T=(\mue^3/3\bar{\mu})^2/2\pi\Delta$ that is
as tiny as $\sim$ eV for $\mue\sim$ a few $\text{MeV}$.  Consequently
there is no gapless $rs$-$bu$ quarks any more already at tiny
temperatures.


\section{potential curvature}

In order to derive the information on the \textit{full} evaluation of
screening masses, let us elaborate a numerical method based on the
previous analytical consideration to compute the Debye and Meissner
masses here.  It is well-known that the Debye mass can be expressed as
the potential curvature as follows~\cite{kapusta};
\begin{equation}
 m_{D,\alpha\beta}^2 = -\frac{\partial^2 \Omega_{\mu}}
  {\partial\mu_\alpha \partial\mu_\beta} \,,
\label{eq:potential_debye}
\end{equation}
where $\mu$'s are the color chemical potential coupled to the
generators $T_\alpha$.  That means, $\mu$'s here involve off-diagonal
components in color-flavor space besides (\ref{eq:mu}).  The
thermodynamic potential defined with all $\mu$'s is denoted by
$\Omega_\mu$ here.  The potential curvature (\ref{eq:potential_debye})
is evaluated at vanishing $\mu$'s except $\mu_3$ and $\mu_8$ which are
fixed by the color neutrality conditions.In QCD at finite temperature,
moreover, pure gluonic loops produce a Debye mass $\sim gT$, which is
not included in Eq.~(\ref{eq:potential_debye}).

\begin{figure}
\includegraphics[width=7cm]{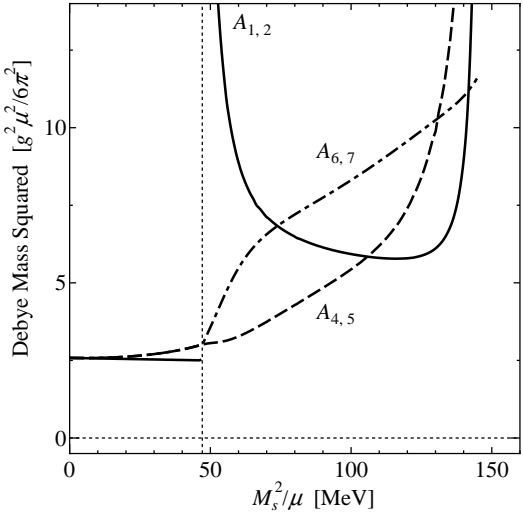}
\caption{Debye mass for $A_{1,2}$ (solid), $A_{4,5}$ (dashed), and
$A_{6,7}$ (dot-dashed) obtained from the potential curvature at
$T=0$.}
\label{fig:debye12}
\end{figure}

\begin{figure}
\includegraphics[width=7cm]{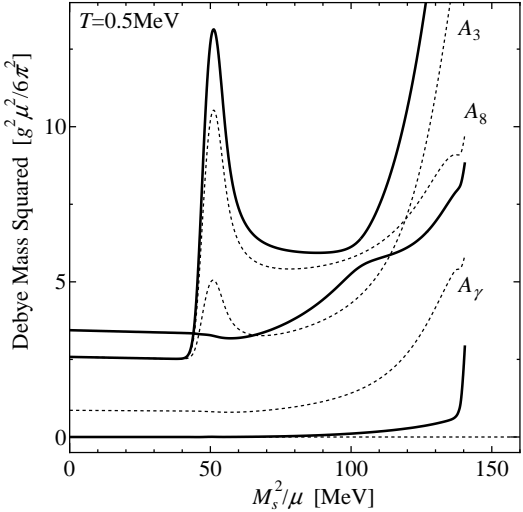}
\caption{Debye mass for the color-diagonal gluons $A_3$, $A_8$ and
photon $A_\gamma$ at $T=0.5\MeV$.  Three solid curves represent three
eigenvalues, and dotted curves are the Debye masses with mixing turned
off by hand.  The electromagnetic coupling constant is chosen to be
$e/g=1/2$, which is presumably larger than a realistic value, to make
visible the mixing effect of $A_\gamma$.}
\label{fig:debye380}
\end{figure}

All we have to know to evaluate the thermodynamic potential are the
energy dispersion relations that are much easier than calculating the
loop integral (\ref{eq:self-energy}) directly.  We show the Debye
masses as a function of $\mssq$ inferred from
(\ref{eq:potential_debye}) in Figs.~\ref{fig:debye12} and
\ref{fig:debye380}.  The singular behavior associated with the gapless
onset emerges in an upward direction, that is accounted for from the
relation (\ref{eq:diag(p-h)}).  In other words, through the relation
(\ref{eq:diag(p-h)}), the positively rising results for the Debye
masses squared imply negative Meissner masses squared for $A_{1,2}$
and $A_{3,8,\gamma}$ around the gCFL onset and for all the gluons for
large $\mssq$.  It is worth noting that the $\Qtilde$-photon has a
finite Debye mass for large $\mssq$ because the system in the gCFL
phase has a nonzero density of electrons that can screen
$\Qtilde$-charge.

Once we develop a relation connecting the thermodynamic potential
to the Meissner mass like the above relation
(\ref{eq:potential_debye}) with respect to the Debye mass, we can
numerically compute the Meissner mass including not only the singular
parts but also all the contributions.  As a matter of fact, we have
found a relation,
\begin{equation}
 (m_{M,\alpha\beta}^2)_{\text{bare}} = \frac{1}{3}\sum_{i=1}^3
  \frac{\partial^2 \Omega_{A}}{\partial A_{\alpha,i}
  \partial A_{\beta,i}}\biggr|_{A=0} \,,
\label{eq:potential_meissner}
\end{equation}
where ``bare'' means the Meissner mass squared with terms diverging
like $\sim\Lambda^2$, and $\Omega_A$ is the thermodynamic potential
defined in the presence of the gauge fields $\boldsymbol{A}_i$ (Note
that bold symbols stand for matrices in color-flavor space in our
notation.) in the Lagrangian (\ref{eq:lagrangian}) as
\begin{equation}
 {\cal L}_A = \bar{\psi}(i\feyn{\partial}
  +\boldsymbol{\mu}_{\text{eff}}\gamma^0 - \vec{\boldsymbol{A}}\cdot
  \vec{\gamma})\psi + {\cal L}_{\text{int}} \,.
\end{equation}
Equation (\ref{eq:potential_meissner}) needs some more explanation.
It should be worth mentioning that (\ref{eq:potential_meissner}) is a
well-known general relation \textit{if we do accomplish the angle
integration with respect to the loop momentum $\boldsymbol{p}$}.
However, this is technically difficult, for $\boldsymbol{p}$ makes an
angle with $\boldsymbol{A}_i$ on evaluating $\Omega_A$.  The benefit
of (\ref{eq:potential_meissner}) is that we can reduce the
$p$-integration into a one-dimensional $p_3$-integration as if the
rotational symmetry were not affected by $\boldsymbol{A}_i$.  We shall
take advantage of this simplification at the price of treating $\ms$
as chemical potential shifts.

This relation can be proved in the following way.  When the momentum
$\boldsymbol{p}$ in the loop-integral is chosen to be in the
$3$-direction, we can show that,
\begin{equation}
 \begin{split}
 & {\cal T}_{\text{diag(p-p)}}^{11}(p;0)
  ={\cal T}_{\text{diag(p-p)}}^{22}(p;0) = 0, \\
 & {\cal T}_{\text{diag(p-p)}}^{33}(p;0) = 2, \\
 & {\cal T}_{\text{diag(p-a)}}^{11}(p;0)
  ={\cal T}_{\text{diag(p-a)}}^{22}(p;0) = 2,\\
 & {\cal T}_{\text{diag(p-a)}}^{33}(p;0) = 0,
 \end{split}
\label{eq:T123}
\end{equation}
indicating that $\Pi^{33}$ contains only the p-p contribution and
$\Pi^{11}$ and $\Pi^{22}$ the p-a contribution.  From these
expressions in the same way as (\ref{eq:calc_meissner}) we eventually
arrive at the relation,
\begin{equation}
 m_{M,\alpha\beta}^2 = \third\bigl[\Pi_{\alpha\beta}^{11}(0,0)
  +\Pi_{\alpha\beta}^{22}(0,0)+\Pi_{\alpha\beta}^{33}(0,0)\bigr] \,.
\label{eq:meissner_polarization}
\end{equation}
We should remark that, even though we do not know an explicit
expression for the $ru$-$gd$-$bs$ quark propagator, the ($\mu,\nu$)
structure comes out only through the energy projection operators and
the above argument is generally valid for any quark propagator as long
as $\ms$ can be treated as an effective chemical potential shift. The
relation (\ref{eq:meissner_polarization}) is readily rewritten in
terms of the thermodynamic potential, which leads us to
(\ref{eq:potential_meissner}).

\begin{figure}
\includegraphics[width=7cm]{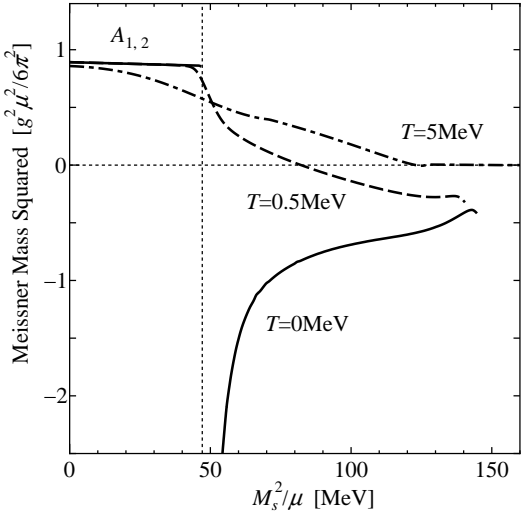}
\caption{Meissner mass squared for $A_{1,2}$ as a function of $\mssq$
at the temperatures $T=0$ (solid), $0.5\MeV$ (dashed), and $5\MeV$
(dot-dashed).  The vertical dotted line represents the gCFL onset.
The model parameters are chosen to yield $\Delta_0=25\MeV$ at zero
temperature and zero strange quark mass.  The solution of the coupled
equations ceases to exist at $\mssq=144.7\MeV$ and $140.5\MeV$ for
$T=0$ and $0.5\MeV$ respectively.}
\label{fig:a12}
\end{figure}

\begin{figure}
\includegraphics[width=7cm]{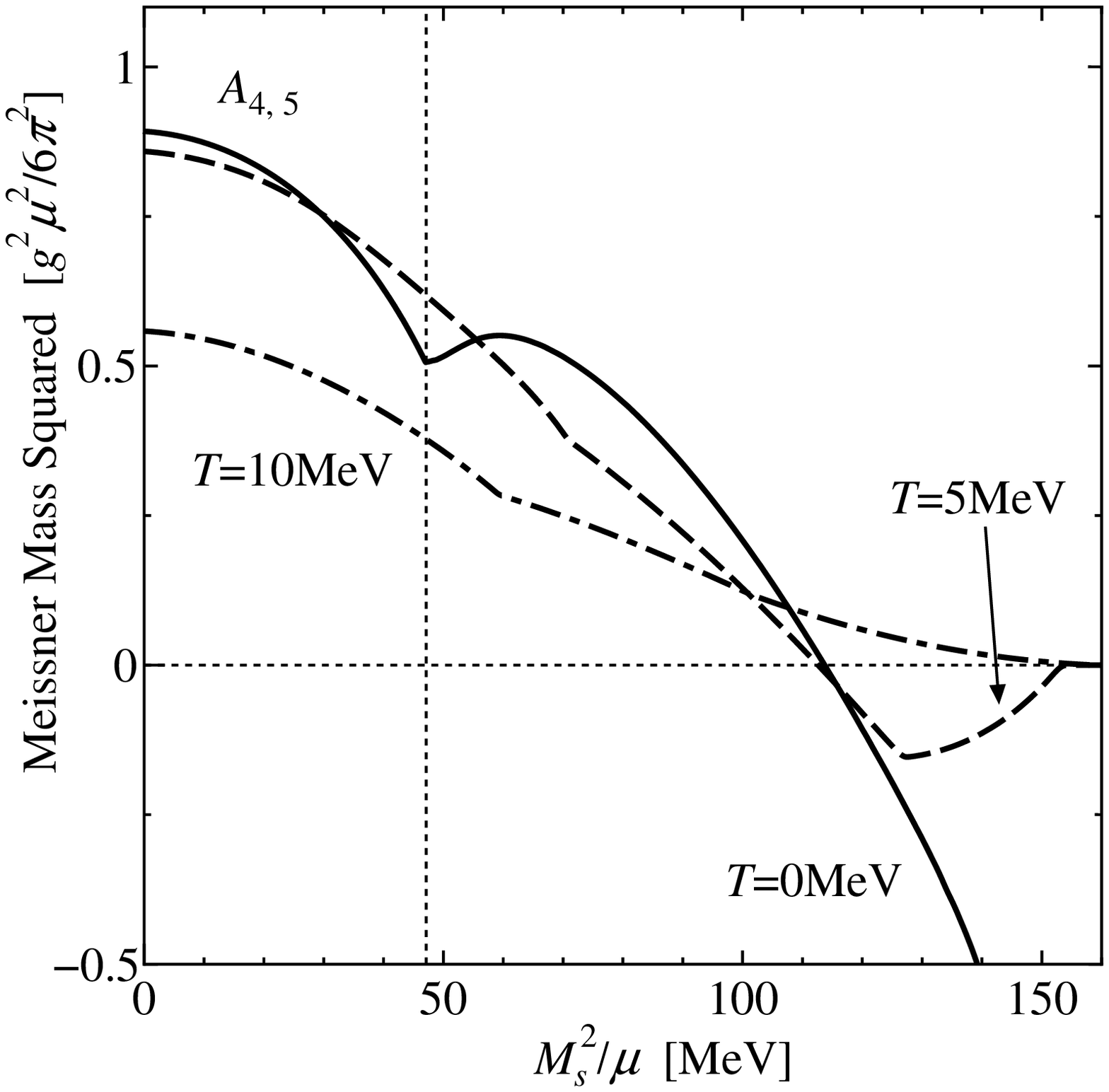}
\caption{Meissner mass squared for $A_{4,5}$ as a function of $\mssq$
at $T=0$ (solid), $5\MeV$ (dashed), and $10\MeV$ (dot-dashed).}
\label{fig:a45}
\end{figure}

\begin{figure}
\includegraphics[width=7cm]{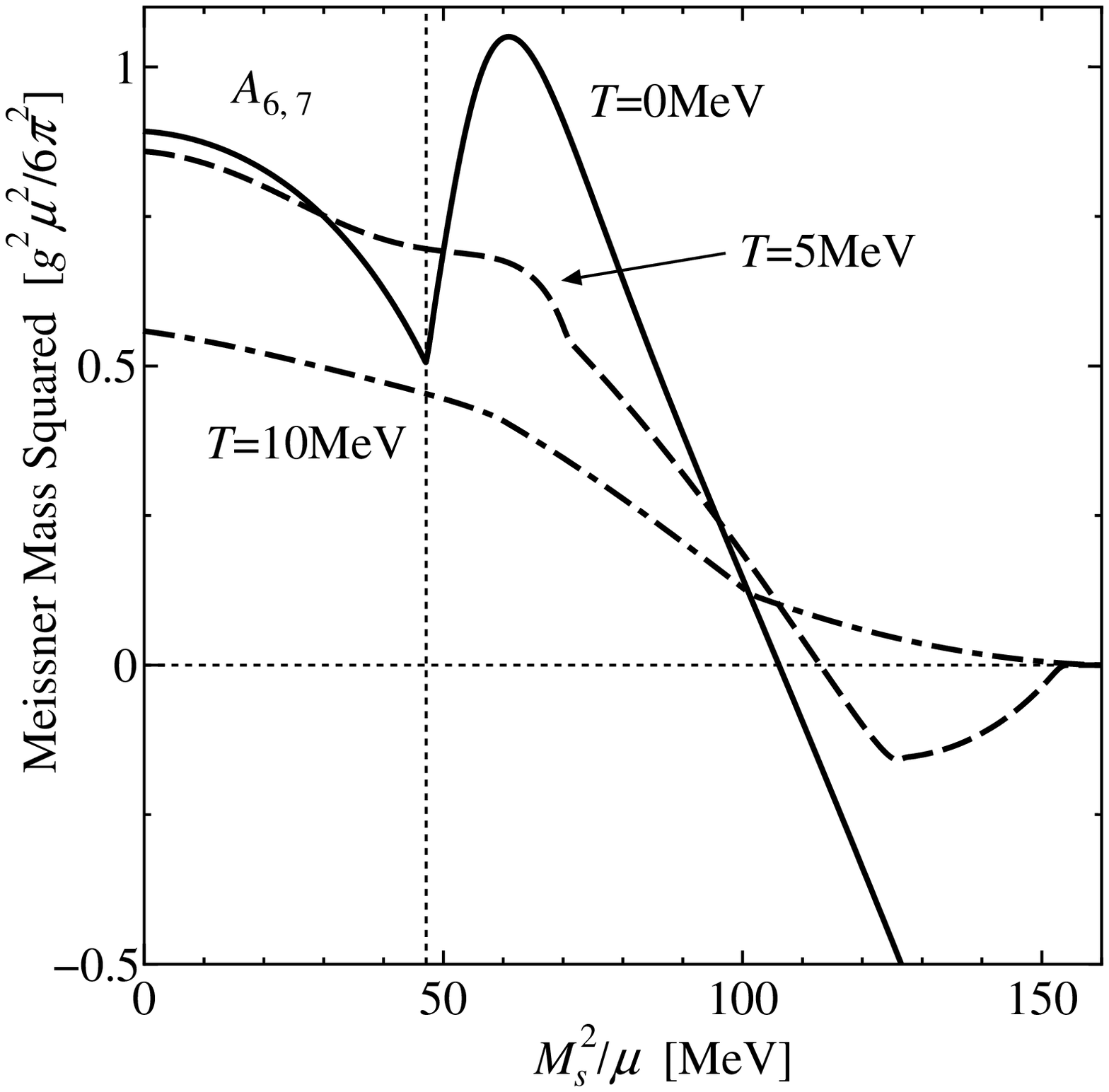}
\caption{Meissner mass squared for $A_{6,7}$ as a function of $\mssq$
at $T=0$ (solid), $5\MeV$ (dashed), and $10\MeV$ (dot-dashed).}
\label{fig:a67}
\end{figure}

\begin{figure}
\includegraphics[width=7cm]{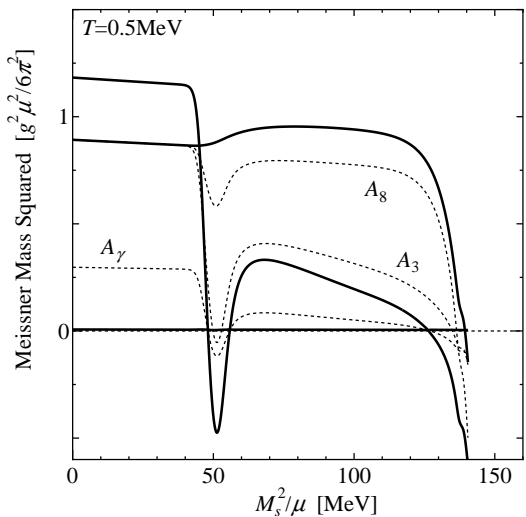}
\caption{Meissner mass squared for three eigenmodes composed of $A_3$,
$A_8$, and $A_\gamma$ as a function of $\mssq$ at $T=0.5\MeV$.  The
electromagnetic coupling constant is chosen to be $e/g=1/2$.  The
dotted curves are for $A_3$, $A_8$, and $A_\gamma$ with mixing among
them turned off.}
\label{fig:a380}
\end{figure}

In our model study, the four-fermion interaction is non-renormalizable
and $\Omega_A$ suffers ultraviolet divergent terms proportional to
$\Lambda^2$.  Thus the Meissner masses derived directly from the
potential curvature should have those divergences to be subtracted
properly by hand.  As argued in the previous analytical study, we
conjecture that the subtraction term is simply inferred by counting
divergent diagrams, which results in;
\begin{align}
 m_{M,\alpha\alpha}^2 &= (m_{M,\alpha\alpha}^2)_{\text{bare}}
  + 3\times \frac{g^2\Lambda^2}{6\pi^2},
  \quad (1\le\alpha\le8)
\label{eq:subtract_3} \\
 m_{M,\gamma\gamma}^2 &= (m_{M,\gamma\gamma}^2)_{\text{bare}}
  + 4\times \frac{e^2\Lambda^2}{6\pi^2},\\
 m_{M,\alpha\beta}^2 &= (m_{M,\alpha\beta}^2)_{\text{bare}}\,,
  \quad (1\le\alpha\neq\beta\le8) \\
 m_{M,\alpha\gamma}^2 &= (m_{M,\alpha\gamma}^2)_{\text{bare}}\,.
\end{align}
These are all determined simply from the diagrammatic counting of the
Nambu-Gor'kov diagonal p-a loops and the associated subtraction
(\ref{eq:subtraction}) with proper coefficients.  In the case of $A_8$
for example, the divergence factor in unit of $\Lambda^2/6\pi^2$
coming from (\ref{eq:sing1}) is
$\half\times(\third+\frac{4}{3})=\frac{5}{6}$ where $\half$ is a
combinatorial factor.  In the same way (\ref{eq:sing2}) and
(\ref{eq:non-singular}) give $\frac{5}{6}$ and $\frac{8}{6}$
respectively.  The sum of these results in $3\times(\Lambda^2/6\pi^2)$
as in (\ref{eq:subtract_3}).  It is interesting to see that the
divergent terms cancel each other in the color-mixing and
photon-mixing channels.  From these above expressions we have obtained
the results as presented in Figs.~\ref{fig:a12}, \ref{fig:a45},
\ref{fig:a67}, and \ref{fig:a380}.

Now, before addressing the instability problem, let us compare our
numerical results shown in Figs.~\ref{fig:a12}, \ref{fig:a45},
\ref{fig:a67}, and \ref{fig:a380} to the known analytical estimates.
In view of our results at $\ms=0$, all the gluons except $A_8$ have
the identical Meissner mass $0.892m_g^2$ which is consistent with
(\ref{eq:Meissner}).  We have confirmed that the agreement becomes
better for larger $\mu\gg\Delta$ because the analytical estimates
require this.  The mixing channel with $e/g=1/2$ (chosen presumably
larger than realistic values to make the mixing effects apparently
visible) results in $1.18m_g^2$ for $\tilde{A}_8$ and zero for
$\tilde{A}_\gamma$, which is also consistent with the analytic
estimate $\tilde{m}_{M,88}^2=1.15m_g^2$ from
(\ref{eq:Meissner_8}). All these suggest that our method works
well.

Figure~\ref{fig:a12} shows the Meissner mass squared for $A_{1,2}$ as
a function of $\mssq$.  The vertical dotted line corresponds to the
gCFL onset and is located near the onset condition given kinematically
$(\mssq)_{\rm c}\simeq2\Delta$. (We chose the model parameters to
yield $\Delta_0=25\MeV$ for $\ms=T=0$.)  We see that the Meissner mass
squared is negative in the entire gCFL region at zero temperature,
that means chromomagnetic instability.  The zero temperature behavior
for $A_{1,2}$ is consistent with the results reported in
Ref.~\cite{Casalbuoni:2004tb} except for the asymptotic property that
the Meissner mass in Fig.~\ref{fig:a12} does not approach zero for
large $\mssq$ as it does in Ref.~\cite{Casalbuoni:2004tb}.  The
singularity around the gCFL onset is drastically smeared by the
temperature effect as seen in the curve for $T=0.5\MeV$, though
chromomagnetic instability still exists for $\mssq>84\MeV$.  At higher
temperatures, as shown by the $T=5\MeV$ curve, the Meissner screening
mass smoothly goes to zero with increasing $\mssq$ since the system
then comes to reach the normal phase without any first-order phase
transition.  (The $T=5\MeV$ results in fact have a tiny unstable
region at $\mssq=125\MeV$, see Fig.~\ref{fig:phase12}.)

In Fig.~\ref{fig:a45} (and in Fig.~\ref{fig:a67}) we plot the results
for $A_{4,5}$ (and for $A_{6,7}$ respectively).  The gross features
are in good agreement with the results presented in
Ref.~\cite{Casalbuoni:2004tb} and all these gluons do not suffer
chromomagnetic instability until $\mssq>105\MeV$ at zero temperature.
We remark that the instabilities for $\mssq\gtrsim100\MeV$ persist
even at $T=5\MeV$, but it disappears at $T=10\MeV$.

The Meissner masses corresponding to $A_3$, $A_8$, and $A_\gamma$, as
we have exposited before, are negatively divergent if any of quark
excitation energies take the quadratic form as a function of the
momentum.  This situation takes place only at the gCFL onset.  The
results inside the gCFL regime are negatively large but not divergent
actually.  In fact a finite density of electrons requires an almost
but not exactly quadratic dispersion relation anywhere in the gCFL
phase.  The outputs accordingly have severe dependence on the precise
form of the almost quadratic dispersion relation, that is, the
deviation from the quadratic form, which is hard to evaluate
numerically with an extreme accuracy.  We have confirmed the above
mentioned behavior quantitatively in our calculations, that is,
negatively large Meissner masses squared for $A_3$, $A_8$, and
$A_\gamma$ in the entire gCFL region, but we could not reduce
tremendous numerical ambiguity.  Instead of tackling numerical
difficulties that would be milder at finite temperature, we have
calculated the Meissner masses at $T=0.5\MeV$.  In Fig.~\ref{fig:a380}
the dotted curves are the masses without mixing taken into account and
the solid curves are the eigenvalues of the $3\times3$ mass squared
matrix.  We clearly see that instability certainly occurs in one or
two of three eigenmodes.  Furthermore, we note that any effect of the
almost quadratic dispersion relation cannot be perceived from the
$T=0.5\MeV$ results because it is already smooth enough at that
temperature as is obvious also in Fig.~\ref{fig:sing}.

\begin{figure}
\includegraphics[width=7cm]{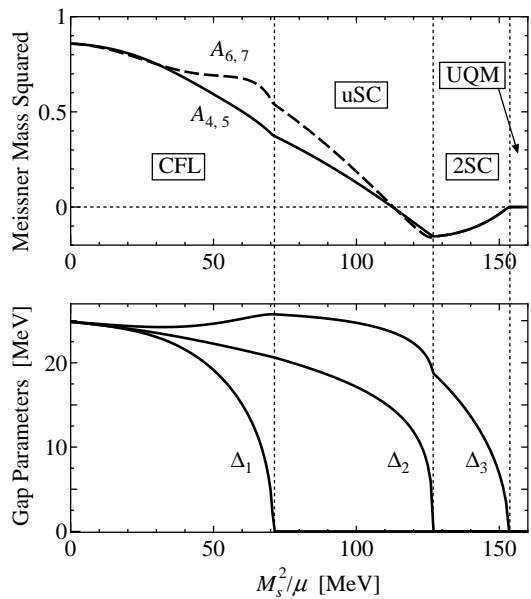}
\caption{The correspondence between chromomagnetic instability for
$A_{4,5}$) and $A_{6,7}$ and the CFL, uSC, and 2SC phases as well as
unpaired quark matter (UQM).  The temperature is chosen to be
$T=5\MeV$ to draw this figure.}
\label{fig:gaps}
\end{figure}

The unstable regions read from these figures are to be considered as
slices with $T$ fixed on the phase diagram.  It is instructive to
compile all these slices and map respective unstable regions onto the
phase diagram.  Figure~\ref{fig:gaps} is one example to locate the
phase boundary and the unstable region and to see them on a common
basis.  In this way we clarified the unstable regions on the phase
diagram as we have already discussed at the beginning of this paper.


\section{summary and open questions}

In summary, in dense neutral three-flavor quark matter, we figured out
the computational procedure to derive the Debye and Meissner screening
masses from curvatures of the thermodynamic potential with respect to
gluon source fields.  We investigated chromomagnetic instability by
changing the temperature $T$ and the Fermi momentum mismatch $\mssq$
and explored the unstable regions for respective gluons and photon
on the phase diagram involving the (g)CFL, uSC, and (g)2SC phases.

\begin{figure}
\includegraphics[width=7cm]{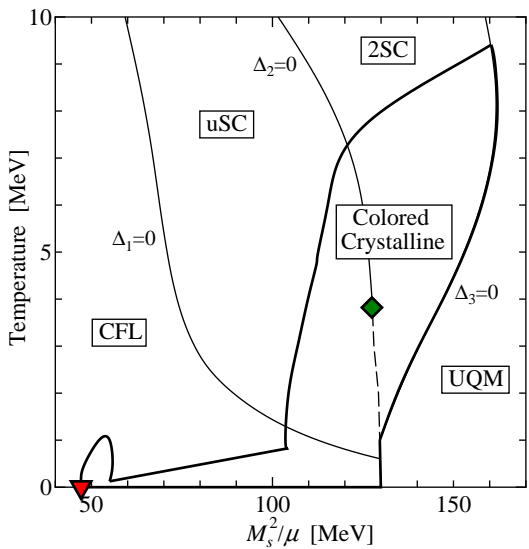}
\caption{The unstable region in which at least one gluon has imaginary
Meissner mass is shown enclosed with solid thick lines, which can be
interpreted as boundaries between the homogeneous and (colored)
crystalline phases.}
\label{fig:phase_all}
\end{figure}

We can say that the instability lines we added on the renewed phase
diagram tell us, at least, at what $\mu$ or $\ms$ the homogeneous
(g)CFL, uSC, and (g)2SC phases remain to be stable at a certain $T$.
From the present work, however, we cannot extract any clue about what
is going on inside the unstable regions.  One possible solution would
be a crystalline-like phase as is suggested in the two-flavor
case~\cite{Giannakis:2004pf}.  If this is true also in three-flavor
quark matter, it is most likely that the instability lines are to be
regarded as the boundaries that separate the homogeneous and
crystalline superconducting phases.  In this sense, the boundaries
drawn by thick lines in Figs.~\ref{fig:phase12}, \ref{fig:phase4567},
and \ref{fig:phase380} have a definite physical meaning even after a
true ground state inside the unstable regions will be identified.
Strictly speaking, once chromomagnetic instability occurs in any gluon
channel, it would affect the other gluon channels.  Consequently
instability lines with respect to one gluon going inside the unstable
region with respect to another gluon might be substantially different
from what we have seen in this work.  They are actually the most outer
boundaries only that separate the homogeneous and crystalline phases,
as shown in Fig.~\ref{fig:phase_all}.

Our approach here is essentially along the same line as in
Ref.~\cite{Fukushima:2004zq}, and so this work leaves the same open
questions enumerated in Ref.~\cite{Fukushima:2004zq}; how gauge field
fluctuations affect the critical point and the order of respective
phase transitions~\cite{Matsuura:2003md},  what the nature of
$K^0$-condensation in the gCFL phase is and how it affects
instability~\cite{Kryjevski:2004jw,Buballa:2004sx},  what difference
the 't~Hooft (instanton) interaction makes that induces an interaction
like $\sim|\Delta_3|^2 \ms$, where the chiral phase transition is
located and how our ($\mssq$-$T$) phase diagram is mapped onto the
($\mu$-$T$) plane with $\ms$ solved dynamically.

In addition to these above mentioned issues, we should make an
improvement how to take account of $\ms$ fully.  The derivation of our
formula (\ref{eq:potential_meissner}) needs the energy projection
operator for massless quarks and works out as long as the $\ms$
effects can be well-approximated by a chemical potential shift.  From
this reason, we could not augment the phase diagrams for stronger
couplings (Figs.~16 and 17 in Ref.~\cite{Fukushima:2004zq} for
example) with the instability lines.  The improvement could be
implemented by the full angle-integration in $\boldsymbol{p}$ to
acquire $\Omega_A$.

From the theoretical point of view, logarithmic divergences that have
been simply neglected in the present work must be taken seriously.  In
particular, logarithmic terms depending on $\ms$ seem to remain even
in the normal phase~\cite{Alford}, while in our work the Meissner
masses turn out to be zero properly in the normal phase, suggesting
some cancellation mechanism between $\ms$ dependent divergences.  This
must be further investigated in a field-theoretical way.  Another
problem of theoretical interest is how much singularities are
inhibited by non-perturbative resummation.  However small the coupling
constant $g$ is, the polarization is divergently large at the one-loop
level, and generally, that demands a sort of resummation, for
instance, by means of the ladder approximation.  The instability might
not be cured itself by resummation, but the singularities near the
gCFL onset would be possibly smoothened.

Phenomenologically, since not only the gCFL but also the g2SC phase is
within our scope especially once we will be able to approach the
stronger coupling case, it is inevitable to examine the competition
between the mixed and crystalline phases around the g2SC region.  The
entanglement between these phase possibilities should play an
important role to complement the phase diagram for the large $\ms$ or
moderate density region~\cite{iida}.

\begin{acknowledgments}
The author acknowledges helpful conversations with M.~Alford, M.~Forbes,
C.~Kouvaris, M.~Huang, K.~Iida, K.~Rajagopal, A.~Schmitt, and
I.~Shovkovy.  This work was supported in part by RIKEN BNL Research
Center and the U.S.\ Department of Energy under cooperative research
agreement \#DE-AC02-98CH10886.
\end{acknowledgments}

\end{document}